# A New Rate Region for General Interference Channel (Improved HK Region)

Ghosheh Abed Hodtani
Department of Electrical Engineering, Ferdowsi University of Mashhad
ghodtani@gmail.com

**Abstract.** In this paper **(a)** after detailed investigation of the previous equivalent rate regions for general interference channel, i.e., the Han-Kobayashi (HK) and the Chong-Motani-Garg (CMG) regions, we define modified CMG region the equivalency of which with the HK region is readily seen; **(b)** we make two novel changes in the HK coding. First, we allow the input auxiliary random variables to be correlated and, second, exploit the powerful technique of random binning instead of the HK -CMG superposition coding, thereby establishing a new rate region for general interference channel, as an improved version of the HK region; **(c)** we make a novel change in the CMG coding by allowing the message variables to be correlated and obtain an equivalent form for our new region in (b), as an improved version of the CMG region. Then, **(d)** in order to exactly demarcate the regions, by considering their different easily comparable versions, we compare our region to the HK and CMG regions. Specifically, using a simple dependency structure for the correlated auxiliary random variables, based on the Wyner and Gacs-Korner common information between dependent variables, we show that the HK and the CMG regions are special cases of our new region.

**Keywords.** Interference channel, Correlated auxiliary random variables, Common information, Superposition coding, Binning scheme

## I. INTRODUCTION

Interference channel (IC) has been the most important and complicated channel for information theory researchers since its initiation by Shannon [1]; and is recently being studied in great detail due to its wide range of potential applications. Here we consider only the two user IC, where each sender communicates with its respective receiver interfering with communication of the other sender-receiver.

The study of the IC was furthered by Ahlswede [2]. Sato [3] obtained various inner and outer bounds by considering the associated multiple access sub-channel in the IC.

Carleial [4] established an improved achievable rate region (with one auxiliary random variable for each sender ) by using sequential decoding and convex hull operation based on the superposition coding of Cover [5].

Han and Kobayashi (KH) [6],[7], generalized Cover's superposition coding used by Carleial [4] to the many variable case; applied jointly or simultaneous decoding strategy instead of sequential decoding in [4] and the time-sharing formulation instead of convex hull operation in [4] for the general IC, thereby establishing the best achievable rate region known to date.

Chong, Motani and Garg [8], by slightly modifying the decoding error definition and reducing the number of independent auxiliary random variables by superposition coding accompanied with special Markovity chains, derived a simplified description for the HK rate region, which we will refer to as the CMG region. In [8],[7], the equivalence of the regions is proved.

With the exception of a few special cases, the capacity region of the IC is not known. The problem of determining the capacity region and even some rate regions has been studied dominantly from the viewpoint of previously investigated special cases of multiple and broadcast sub-channels in the IC: [6],[9],and [10]-[20]. The Gaussian IC has been intensively studied in [6],[10],[21]-[27],[28].

**Motivations**

Our motivations for this work have been the following.

- The anonymous reviewer of the HK seminal paper [6] has suggested to allow the input auxiliary random variables to be correlated for possibly improving the IC rate region.
- Correlated variables have common information and their transmission is of practical importance and has a communication complexity [29]-[34].For example, from viewpoint of our interest, the mutual information $I(U,W;Y)$ has different values for dependent and independent $(U,W)$.
- Random binning is a powerful technique for coding of dependent variables and increasing the rate [35].
- Mathematically speaking, multiple access channels have been studied, first for independent input variables [36],[37], second for specially correlated [38] and then for arbitrarily correlated ones [39]. In [6], the IC has been studied for independent input auxiliary random variables, and in [7],[8] superposition coding, as a special case of the HK scheme, has been applied on these independent variables.

Therefore, the study of the IC for:

dependent and correlated variables with random binning technique and superposition coding,

is a new issue and also, may explain and demarcate the regions more exactly.

**Our work**

By taking into account the above points and specifically two common similarities of the HK-CMG coding, i.e.,the independency of message variables and superposition coding, and also, generally accepted fact of the equivalency of the HK and the CMG regions:

**(a)** having reviewed the HK and CMG regions in full detail, we define modified CMG region the equivalency of which with the HK region is readily seen;

**(b)** we make two novel changes in the HK –CMG coding strategy. First, we allow the input auxiliary random variables to be correlated or consider a general input distribution and, second, exploit the powerful technique of random binning [40] as in [41], instead of the HK-CMG superposition coding. Then, by using of the HK jointly decoding scheme, we obtain a new rate region for general IC. Then,

**(c)** in the CMG coding, we allow the message variables to be correlated and obtain an equivalent form for our new region in (b) as an improved version of the CMG region. Finally,

**(d)** we exactly demarcate the regions, by considering their different easily comparable versions. Specifically, in view of the latest studies on the Wyner and Gacs-Korner common information [42] between dependent random variables, we use a simple dependency structure and show that, term by term, our region is an improved version of the previous regions (the HK and CMG equivalent regions).

**Paper organization**

The remainder of the paper is as follows. In section II, we define the IC and the modified IC. In section III, we explain the common information between dependent variables and review some points regarding separately coding, superposition coding and random binning scheme for point to point and multiple access sub-channels of the IC. In section IV and V, we have a fully detailed investigation of the HK and the CMG, modified CMG regions, respectively. In section VI, we establish our new rate region (improved HK region) for general IC accompanying with deriving its different versions. In section VII, we derive a simplified description for our region as an improved CMG region. Section VIII consists of detailed comparison of the regions. Finally, we have a conclusion in section IX.

## II. BASIC DEFINITIONS

We denote random variables by $X_1, X_2, Y_1, \cdots$ with values $x_1, x_2, y_1, \cdots$ in finite sets $\mathcal{X}_1, \mathcal{X}_2, \mathcal{Y}_1, \cdots$ respectively; n-tuple vectors of $X_1, X_2, Y_1, \cdots$ are denoted with $\boldsymbol{x_1}, \boldsymbol{x_2}, \boldsymbol{y_1}, \cdots$. We use the symbol $A_\varepsilon^n(X_1, X_2, \cdots, X_l)$ to indicate the set of $\varepsilon$-typical n-sequences $(\boldsymbol{x_1}, \boldsymbol{x_2}, \cdots, \boldsymbol{x_l})$ [43].

**Interference Channel (IC)**

A discrete and memoryless IC $(\mathcal{X}_1 \times \mathcal{X}_2, p(y_1 y_2|x_1 x_2), \mathcal{Y}_1 \times \mathcal{Y}_2)$ consists of two sender-receiver pairs ($X_1 \longrightarrow Y_1$ and $X_2 \longrightarrow Y_2$) in Fig.1, where $\mathcal{X}_1, \mathcal{X}_2$ are two finite input alphabet sets; $\mathcal{Y}_1, \mathcal{Y}_2$ are two finite output alphabet sets, and $p(y_1 y_2|x_1 x_2)$ is a conditional channel probability of $(y_1, y_2) \in \mathcal{Y}_1 \times \mathcal{Y}_2$ given $(x_1, x_2) \in \mathcal{X}_1 \times \mathcal{X}_2$. The nth extension of the channel is:

$$p(\boldsymbol{y_1 y_2}|\boldsymbol{x_1 x_2}) = \prod_{i=1}^{n} p(y_{1i} y_{2i}|x_{1i} x_{2i})$$

A code $(n, M_1 = \lfloor 2^{nR_1} \rfloor, M_2 = \lfloor 2^{nR_2} \rfloor, \varepsilon)$ is a collection of $M_1$ codewords $\boldsymbol{x_{1i}} \in \mathcal{X}_1^n, i \in \mathcal{M}_1$; $M_2$ codewords $\boldsymbol{x_{2j}} \in \mathcal{X}_2^n, j \in \mathcal{M}_2$; two decoding functions $g_1: \boldsymbol{y_1} \longrightarrow \mathcal{M}_1$, $g_2: \boldsymbol{y_2} \longrightarrow \mathcal{M}_2$; and the average error probabilities at the receivers $(P_{e_1}^n, P_{e_2}^n)$ are defined conveniently [6],[8].

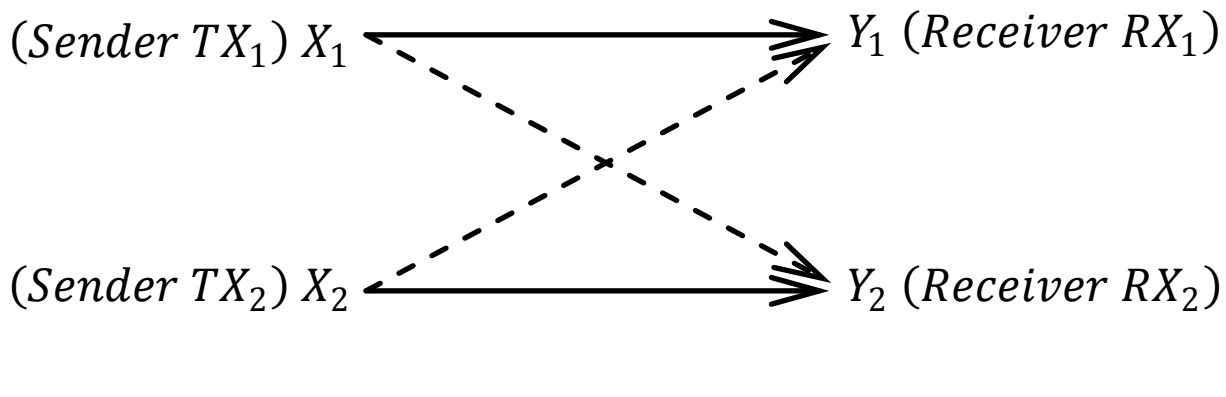

Fig.1 Interference channel

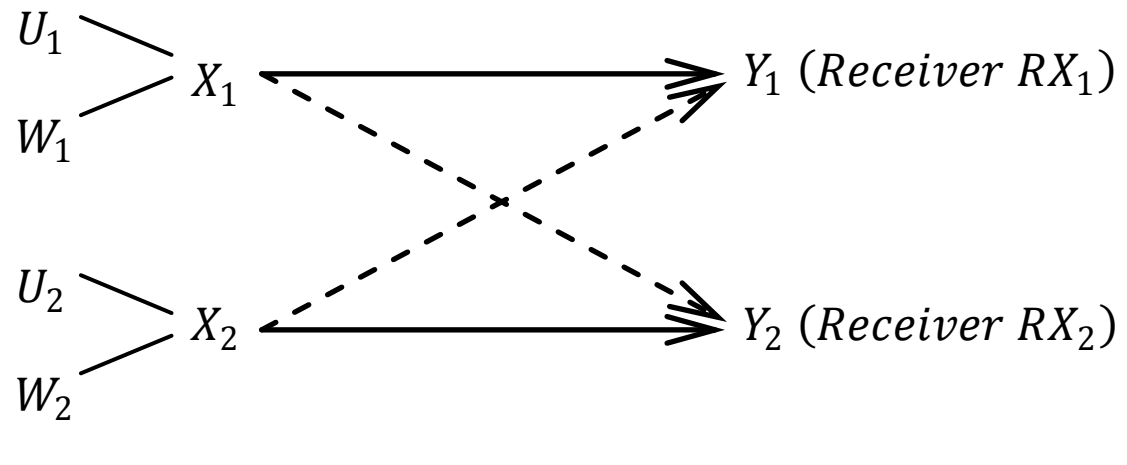

Fig.2 Modified interference channel

A pair $(R_1, R_2)$ of non-negative real values is called an achievable rate if there exists a sequence of codes such that under some decoding scheme, $\max(P_{e_1}^n, P_{e_2}^n) < \varepsilon$.

The capacity region of the IC is the set of all achievable rates.

**Modified interference channel**

As in [6], a modified IC (Fig.2), models two senders communicating both private and common message to two receivers; where the information conveying role of the channel inputs $X_1, X_2$ is transferred to some fictitious inputs $U_1, W_1, U_2, W_2$, so that the channel behaves like a channel $U_1, W_1, U_2, W_2 \longrightarrow Y_1Y_2$.

Auxiliary random variables $W_1$ and $W_2$ represent the public message to be sent from $TX_1$ to $(RX_1, RX_2)$ with the rate $T_1$ and from $TX_2$ to $(RX_1, RX_2)$ with the rate $T_2$, respectively. Similarly, $U_1$ and $U_2$ are the private message to be sent from $TX_1$ to $RX_1$ with the rate $S_1$ and from $TX_2$ to $RX_2$ with the rate $S_2$, respectively. Also, as in [6], $Q \in \mathcal{Q}$ is a time sharing random variable whose n-sequences $\boldsymbol{q} = (q_1, q_2, \cdots, q_n)$ are generated independently of the messages. The n-sequences $\boldsymbol{q}$ are given to both senders and receivers.

An $(n, \lfloor 2^{nT_1} \rfloor, \lfloor 2^{nS_1} \rfloor, \lfloor 2^{nT_2} \rfloor, \lfloor 2^{nS_2} \rfloor, \varepsilon)$ code for the modified IC (Fig.2) consists of $\lfloor 2^{nT_1} \rfloor$ codewords $\boldsymbol{w_1}(j)$, $\lfloor 2^{nS_1} \rfloor$ codewords $\boldsymbol{u_1}(l)$ for $TX_1$; and $\lfloor 2^{nT_2} \rfloor$ codewords $\boldsymbol{w_2}(m)$, $\lfloor 2^{nS_2} \rfloor$ codewords $\boldsymbol{u_2}(k)$ for $TX_2$ ; $j \in \{1, \cdots, 2^{nT_1}\}$, $l \in \{1, \cdots, 2^{nS_1}\}$, $m \in \{1, \cdots, 2^{nT_2}\}$, $k \in \{1, \cdots, 2^{nS_2}\}$, such that the maximum of the average probabilities of decoding error $(P_{e_1}^n, P_{e_2}^n)$ is less than $\varepsilon$.

A quadruple $(T_1, S_1, T_2, S_2)$ of non-negative real numbers is achievable for the modified IC (and hence, $(R_1 = S_1 + T_1, R_2 = S_2 + T_2)$ is achievable rate for the IC) if there exists a sequence of codes such that the maximum of average error probabilities under some decoding scheme is less than $\varepsilon$. An achievable region for the modified IC is the closure of a subset of the positive region $R^4$ of achievable rate quadruples $(T_1, S_1, T_2, S_2)$.

Therefore, we can consider auxiliary random variables $Q, U_1, W_1, U_2, W_2$, defined on arbitrary finite sets $\mathcal{Q}, \mathcal{U}_1, \mathcal{W}_1, \mathcal{U}_2, \mathcal{W}_2$, respectively; $X_1$ and $X_2$ defined on the input alphabet sets $\mathcal{X}_1, \mathcal{X}_2$, and $Y_1, Y_2$, defined on the output alphabet sets $\mathcal{Y}_1$ and $\mathcal{Y}_2$. Let $Z = (QU_1W_1U_2W_2X_1X_2Y_1Y_2)$ and let $\mathcal{P}_{IC}$ be the set of all distributions of the form (for Fig.2): (hereafter, for brevity, let $p(qu_1w_1u_2w_2x_1x_2y_1y_2) = p(\,)$)

$$p(\,) = p(q)p(u_1w_1|q)p(u_2w_2|q)p(x_1|qu_1w_1)p(x_2|qu_2w_2)p(y_1y_2|x_1x_2) \qquad (1).$$

## III. EXPLANATORY PRELIMINARIES

In this section, firstly, we review the important concept of common information; secondly we show that for a point to point channel with two independent message variables, separately coding and superposition coding both result to the same rates, and random binning for dependent message variables improves the rate. Then, the same claim is pointed out for multiple access channels.

### a. Common information

Here, we exemplify the concept of common information between dependent and independent variables, aiming at our interest to use the difference between mutual information $I(Y; U, W)$ and $I(U_dW_d; Y)$, where $U, W$ and $U_d, W_d$ are independent and dependent variables, respectively.

Finding the extent of common information between dependent random variables has received much attention [29]-[34],[42] due to its potential applications in information theory and other research areas.

Suppose that $U_d = (U, K)$ and $W_d = (W, K)$ where $U, W, K$ are independent. In view of Wyner's definition [30], a natural measure of common information between $U_d$ and $W_d$ is the entropy of common part $H(K)$; and conditioned on $K$, there is no residual information or $I(U_d; W_d|K) = 0$. Gacs-Korner common information is a generalized version and is the largest $H(K)$ for which the random variables $U_d = (U, K)$ , $W_d = (W, K)$ consist of possibly dependent $U, W, K$. A generalization of Gacs-Korner has been presented in [44],[42].

Here, we consider a simple dependency structure between $U_d = (U, K)$ and $W_d = (W, K)$, assuming that $U, W, K$ are independent and in an example, we compute $H(K), H(W_d), H(U_d), H(W_d, U_d), H(U, W), I(UW; Y)$ for $Y = U + W$, and $I(U_dW_d; Y)$ for $Y = U_d + W_d$.

**Example**: Suppose that $U_d = (U, K)$ , $W_d = (W, K)$ with alphabets $K = \{0,1\}$, $\mathcal{U} = \mathcal{W} = \{0,1,2,3\}$ and probability distribution as follows.

$$p(u_d, w_d) = \begin{cases} \frac{1}{32} \text{ , } u_d \text{ and or } w_d = (0,0),(1,0),(2,0),(3,0) \text{ and } u_d \text{ and or } w_d = (0,1),(1,1),(2,1),(3,1) \\ 0 \qquad\qquad\qquad\qquad\qquad\qquad \text{, oth.} \end{cases}$$

Then,

$$P(K = 0) = \sum_{W,U} P(K = 0, W, U) = \frac{1}{32} \times 16 = \frac{1}{2}$$

$$P(K = 1) = \sum_{W,U} P(K = 1, W, U) = \frac{1}{32} \times 16 = \frac{1}{2}$$

$$P(W_d) = \sum_{U_d} P(W_d, U_d) \text{ , } P(U_d) = \sum_{W_d} P(U_d, W_d) \text{ , } P(W, U) = \sum_{K} P(W_d, U_d) \text{ ,}$$

$$P(W) = \sum_{U} P(U, W) \text{ , } P(U) = \sum_{W} P(U, W).$$

We observe that in this example, $(W_d, U_d)$ are dependent and the corresponding variables $(U, W, K)$ are independent. We have:

$P(U, W, K) = P(U)P(W)P(K)$ and $H(K) = 1$ , $H(W_d) = H(U_d) = 3$ , $H(W_d, U_d) = -32 \times \frac{1}{32} log \frac{1}{32} = 5 \neq H(W_d) + H(U_d)$ , $H(U, W) = 4$ , $H(U) = 2$ , $H(W) = 2$ , $H(W_d, U_d) = H(W) + H(U) + H(K)$ , $H(W_d) = H(W) + H(K)$ , $H(U_d) = H(U) + H(K)$

And, for $Y = U + W$ , $I(U, W; Y) = H(Y)$ , $Y = 0,1,2,3,4,5,6$ with $(Y = 0) = P(Y = 6) = \frac{1}{16}$ , $P(Y = 1) = P(Y = 5) = \frac{1}{8}$ , $P(Y = 2) = P(Y = 4) = \frac{3}{16}$ , $P(Y = 3) = \frac{1}{4}$ ,

then, $I(Y; U, W) = 2/69$

And for $Y = U_d + W_d = (U + W, 2K)$, we have $I(U_d W_d; Y) = \cdots = 3/69 > I(UW; Y) = 2/69$

And, $I(W_d; U_d) = H(U_d) + H(W_d) - H(W_d, U_d) = 3 + 3 - 5 = 1$ or $I(W_d; U_d) = H(U_d) - H(U_d | W_d) = 3 - 2 = 1$ and $I(W_d; U_d) = H(K) = 1$

**A simple lemma**

For $U_d = (U, K)$ , $W_d = (W, K)$ with independent $U, W, K$ and for any arbitrary $Y$ and for any $Z$ independent of $U, W, K$ , we have:

$$\begin{cases} I(U_d W_d; Y) \geq I(UW; Y) & \text{(b-1)} \\ I(U_d W_d; Y|Z) \geq I(UW; Y|Z) & \text{(b-2)} \\ I(U_d; Y|W_d Z) \geq I(U; Y|WZ) & \text{(b-3)} \\ I(U_d Z; Y|W_d) \geq I(UZ; Y|W) & \text{(b-4)} \\ I(Z; Y|U_d W_d) \geq I(Y; Z|UW) & \text{(b-5)} \end{cases}$$

**Proof.** In view of $U_d = (U, K)$ , $W_d = (W, K)$ with independent $U, W, K$ and $I(X; Y) \leq I(X; Y|Z)$ for independent $X, Z$, the inequalities are obvious.

b. **Coding schemes for IC virtual sub-channel** $(UW)X \longrightarrow Y$

We explain and show that for the channel with two message variables, two coding techniques, i.e., separately coding and superposition coding, both lead to the same rate, and binning scheme improves the rate for dependent message variables.

Let us consider the sub-channel$(U_1 W_1)X_1 \longrightarrow Y_1$ or $(U_2 W_2)X_2 \longrightarrow Y_2$, and generally $(UW)X \longrightarrow Y$ where the message auxiliary variables are described by $U, W$ with the rates $S, T$, respectively, and sent through $X$. We obtain the achievable rates in two cases, (i) $U, W$ are independent variables, (ii) $U, W$ are dependent variables and considered as $U_d = (U, K)$ , $W_d = (W, K)$, with independent $U, W, K$.

**(i-1)** Separately coding and jointly decoding for independent $U, W$

$U, W$ are independent and are mapped to signal space through deterministic function $x = f(uw|q)$ with time sharing variable $q$ and hence, we have a modified channel $UW \longrightarrow Y$, with distribution

$$p(qwuxy) = p(q)p(u|q)p(w|q)p_q(y|uw) , \quad p(x|uwq) \in \{0,1\}, \; p_q(y|uw) = p(y|x) \quad \text{(c).}$$

Using separately coding and jointly decoding, it is easily proved that the following rate (d-1,2,3) is achievable:

$$\begin{cases} S \leq I(U; Y|WQ) & \text{(d-1)} \\ T \leq I(W; Y|UQ) & \text{(d-2)} \\ S + T \leq I(WU; Y|Q) & \text{(d-3)} \end{cases}$$

**(i-2)** Superposition coding and jointly decoding for independent $U, W$

In this case, for every distribution $p(qwux)$ with independent $(w, u)$ we examine the distribution

$$\begin{cases} p(qwx) = \sum_u p(qwux) & \\ p(qwxy) = p(q)p(w|q)p(x|wq)p(y|x) & \text{(e)} \\ \text{where} \quad W \longrightarrow XQ \longrightarrow Y & \text{(f)} \end{cases}$$

According to (e), the message conveyed by $u$ in (c) is superimposed on $w$ through $x$. Using superposition coding and jointly decoding, it can be proved that the following rate is achievable:

$$\begin{cases} S \leq I(Y; X|WQ) & \text{(g-1)} \\ S + T \leq I(Y; X|Q) & \text{(g-2)} \end{cases}$$

Therefore, the rate $R = S + T$ is $I(UW; Y|Q)$ in (d-3) for the case (i-1) and is $I(Y; X|Q)$ in (g-2) for the case (i-2) and from $x = f(uw|q)$ and (f), we have

$$I(Y; X|Q) = I(UW; Y|Q) \quad \text{(h)}$$

,i.e., the two strategies for independent $U, W$ have the same rates.

**(ii)** Random binning and jointly decoding for dependent $U_d W_d$

$$p(qw_d u_d xy) = p(q)p(u_d|qw_d)p(w_d|q)p_q(y|u_d w_d) , \quad p(x|u_d w_d q) \in \{0,1\}, \; p_q(y|u_d w_d) = p(y|x) \quad \text{(j),}$$

the achievable rate of the channel, using random binning and jointly decoding, can be proved to be the following rate :

$$\begin{cases} S \leq I(Y; U_d | Q W_d) & \text{(i-1)} \\ T \leq I(U_d; W_d | Q) + I(Y; W_d | Q U_d) & \text{(i-2)} \\ S + T \leq I(Y; W_d U_d | Q) & \text{(i-3)} \end{cases}$$

**Note:** The rates in (i-1,2,3) are obviously larger than the rates in (g-1,2) or (d-1,2,3), in accordance with the above simple lemma (in the subsection common information), for a simple dependency structure $U_d = (U, K), W_d = (W, K)$.

**c. Coding schemes for IC virtual multiple access sub-channel**

In Fig. 3, we have virtual multiple access sub-channels of IC. Suppose that $U_1, W_1, W_2$ have the rates $S_1, T_1, T_2$ . Then, we can prove that with separately coding of independent $U_1, W_1, W_2$ with code-words $u_1, w_1, w_2$ ; $x_1 = f_1(u_1, w_1), x_2 = f_2(w_2)$ and jointly decoding at the receiver $Y_1$ , the rate region consisting of seven inequalities (3-1,..7 in the HK region in section IV, Theorem 1) is achievable. However, using superposition coding with code-words $w_1, x_1, w_2, x_2$ and jointly decoding at the receiver $Y_1$ results to a region constrained with four inequalities (12-1,..4 in the CMG region in section V, Theorem 6). The two regions are equivalent. The similar regions are achievable for the second virtual sub-channel with receiver $Y_2$. The intersection of corresponding regions for each coding scheme gives the rate region for the IC.

Assumption of dependency for $U_1, W_1, W_2$ and using random binning technique lead to the sub-region (17-1,..7) in section VI, Theorem 14.

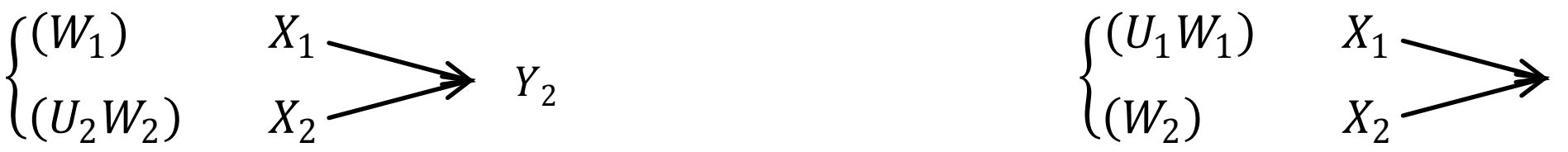

Fig. 3 Multiple access sub-channels of IC in Fig. 2

## IV.THE HK RATE REGION

**The HK input distribution**

Han and Kobayashi [6] considered input auxiliary random variables to be independent and the general distribution (1) in the following form:

$$p(\,) = p(q)p(w_1|q)p(u_1|q)p(w_2|q)p(u_2|q)p_q(y_1 y_2 | u_1 w_1 u_2 w_2) \qquad \text{(2-a)},$$

where $p_q(y_1 y_2 | u_1 w_1 u_2 w_2) = p\big(y_1 y_2 \big| x_1 = f_1(u_1 w_1 | q), x_2 = f_2(u_2 w_2 | q)\big)$, $p(x_i | u_i w_i q) \in \{0,1\}$, $i = 1,2$ and the input distribution is :

$$p(q u_1 w_1 u_2 w_2 x_1 x_2) = p(q)p(w_1|q)p(u_1|q)p(w_2|q)p(u_2|q)\; p(x_1|q w_1 u_1)\; p(x_2|q w_2 u_2) \qquad \text{(2-b)}$$

By using superposition coding of $w_1, u_1, w_2, u_2$ over $q$, with the rates $T_1, S_1, T_2, S_2$ , respectively, and jointly decoding strategy, they derived the best achievable rate region known to date as follows. In the HK coding, the common and private message code-words $(q, u_1, w_1), (q, u_2, w_2)$ are mapped into signal spaces $(x_1), (x_2)$, through arbitrary deterministic functions $f_1, f_2$ respectively, [6].

**The HK region in terms of $(T_1, S_1, T_2, S_2)$**

**Theorem 1** ([6], Theorem 3.1): For the modified IC (Fig.2), let $Z = (Q U_1 W_1 U_2 W_2 X_1 X_2 Y_1 Y_2)$ and let $\mathcal{P}_{IC}^{HK}$ be the set of all distributions of the special form (2-a,b). For any $Z \in \mathcal{P}_{IC}^{HK}$ let $S_{IC}^{HK}(Z)$ be the set of all quadruples $(T_1, S_1, T_2, S_2)$ of non-negative real numbers such that

$$S_1 \leq I(Y_1; U_1 | W_1 W_2 Q) = a_1 \qquad \text{(3-1)},$$
$$T_1 \leq I(Y_1; W_1 | U_1 W_2 Q) = b_1 \qquad \text{(3-2)},$$
$$T_2 \leq I(Y_1; W_2 | U_1 W_1 Q) = c_1 \qquad \text{(3-3)},$$
$$S_1 + T_1 \leq I(Y_1; U_1 W_1 | W_2 Q) = d_1 \qquad \text{(3-4)},$$
$$S_1 + T_2 \leq I(Y_1; U_1 W_2 | W_1 Q) = e_1 \qquad \text{(3-5)},$$
$$T_1 + T_2 \leq I(Y_1; W_1 W_2 | U_1 Q) = f_1 \qquad \text{(3-6)},$$
$$S_1 + T_1 + T_2 \leq I(Y_1; U_1 W_1 W_2 | Q) = g_1 \qquad \text{(3-7)},$$
$$S_2 \leq I(Y_2; U_2 | W_1 W_2 Q) = a_2 \qquad \text{(3-8)},$$
$$T_2 \leq I(Y_2; W_2 | U_2 W_1 Q) = b_2 \qquad \text{(3-9)},$$
$$T_1 \leq I(Y_2; W_1 | U_2 W_2 Q) = c_2 \qquad \text{(3-10)},$$
$$S_2 + T_2 \leq I(Y_2; U_2 W_2 | W_1 Q) = d_2 \qquad \text{(3-11)},$$
$$S_2 + T_1 \leq I(Y_2; U_2 W_1 | W_2 Q) = e_2 \qquad \text{(3-12)},$$
$$T_1 + T_2 \leq I(Y_2; W_1 W_2 | U_2 Q) = f_2 \qquad \text{(3-13)},$$
$$S_2 + T_1 + T_2 \leq I(Y_2; U_2 W_1 W_2 | Q) = g_2 \qquad \text{(3-14)},$$

then any element of the closure of $\bigcup_{Z\in\mathcal{P}_{IC}^{HK}} S_{IC}^{HK}(Z)$ is achievable.

**Proof.** Refer to [6].

**Note.** Hereafter, $a_i, b_i, c_i, d_i, e_i, f_i$ and $g_i$, $i = 1,2$ are the same as in Theorem 1, unless otherwise stated.

**Polymatroidal inequalities for the HK region**

It is easy to verify that the following polymatroidal inequalities between the bound constants $a_i, b_i, c_i, d_i, e_i, f_i$ and $g_i$, $i = 1,2$ hold [7] :

$$\begin{cases} a_i \le d_i \le a_i + b_i \\ \qquad b_i \le d_i \end{cases} \quad , \ i = 1,2$$
$$\begin{cases} a_i \le e_i \le a_i + c_i \\ \qquad c_i \le e_i \end{cases} \quad , \ i = 1,2 \qquad \text{(HK-ineq-1,2,…,30)}$$
$$\{ c_i, b_i \le f_i \le b_i + c_i \quad , \ i = 1,2$$
$$\begin{cases} e_i, d_i \le g_i \le c_i + d_i \ , b_i + e_i \ , a_i + f_i \\ \qquad f_i \le g_i \end{cases} \quad , \ i = 1,2$$

**The HK region in terms of ($R_1$ , $R_2$)**

Now, we transform the above region into the rate pair ($R_1 = S_1 + T_1$ , $R_2 = S_2 + T_2$) using the Fourier-Motzkin elimination technique accompanying with polymatroidal inequalities (HK-ineq-1,2,…,30) and apply the independence of $U_i$ and $W_i$ given $Q$, $i = 1,2$ in the distribution (2-b) to the results and amend Theorem B in [7].

**Theorem 2 [** Theorem B in [7] ]: The region in Theorem 1 can be described as $\mathcal{R}_{HK}$ being the set of ($R_1$ , $R_2$) satisfying:

$R_1 \le d_1$ (4-1),

$\boldsymbol{R_1 \le a_1 + c_2}$ (4-2),

$R_2 \le d_2$ (4-3),

$\boldsymbol{R_2 \le a_2 + c_1}$ (4-4),

$R_1 + R_2 \le a_1 + g_2$ (4-5),

$R_1 + R_2 \le a_2 + g_1$ (4-6),

$R_1 + R_2 \le e_1 + e_2$ (4-7),

$2R_1 + R_2 \le a_1 + g_1 + e_2$ (4-8),

$2R_2 + R_1 \le a_2 + g_2 + e_1$ (4-9),

where $a_i, b_i, c_i, d_i, e_i, f_i$ and $g_i$, $i = 1,2$ are the same as in Theorem 1.

**Proof.** Refer to the proof of Theorem B in [7]. However, in theorem B [7] there are two additional inequalities:

$2R_1 + R_2 \le 2a_1 + e_2 + f_2$ (4-10)

$2R_2 + R_1 \le 2a_2 + e_1 + f_1$ (4-11).

The inequalities (4-10,11) are obtained from (4-2,5) and (4-4,6), respectively, as follows and hence are redundant, merely as a result of the independence of $U_i$ and $W_i$ given $Q$, $i = 1,2$ in (2-b):

$$(4\text{-}2) + (4\text{-}5) \Longrightarrow 2R_1 + R_2 \le 2a_1 + c_2 + g_2 \qquad (5),$$

and the independence of $U_2$ and $W_2$ given $Q$ results in

$$\mathrm{I}(\mathrm{Y}_2;\ \mathrm{U}_2|\mathrm{Q}) \le \mathrm{I}(\mathrm{Y}_2;\ \mathrm{U}_2|\mathrm{QW}_2) \qquad (6),$$

from where we have:

$$\mathrm{c}_2 + \mathrm{g}_2 = \mathrm{I}(\mathrm{Y}_2;\ \mathrm{W}_1|\mathrm{W}_2\mathrm{U}_2\mathrm{Q}) + \mathrm{I}(\mathrm{Y}_2;\ \mathrm{U}_2\mathrm{W}_1\mathrm{W}_2|\mathrm{Q}) \le \mathrm{e}_2 + \mathrm{f}_2 = \mathrm{I}(\mathrm{Y}_2;\ \mathrm{U}_2\mathrm{W}_1|\mathrm{W}_2\mathrm{Q}) + \mathrm{I}(\mathrm{Y}_2;\ \mathrm{W}_1\mathrm{W}_2|\mathrm{QU}_2) \qquad (7).$$

Therefore, in accordance with (7), the relation (5) yields (4-10), i.e. ,(4-10) is redundant. Similarly, (4-4,6) result in the redundancy of (4-11).

Note that the fact we have considered in Theorem 2 is the intrinsic independence of $U_i$ and $W_i$ given $Q$, $i = 1,2$ in the HK region**.**

**An easily comparable form for the ($R_1$ , $R_2$) HK region**

**Theorem 3.** Explanatory and easily comparable form of $\mathcal{R}_{HK}$ in Theorem 2 can be described as $\mathcal{R}_{HK-equi.}$ satisfying thirteen relations for ($R_1$ , $R_2$).

$R_1 \le d_1$ (4-1),

$\boldsymbol{R_1 \le a_1 + c_2}$ (4-2),

$R_2 \le d_2$ (4-3),

$\boldsymbol{R_2 \le a_2 + c_1}$ (4-4),

$R_1 + R_2 \le a_1 + g_2$ (4-5),

$R_1 + R_2 \le a_2 + g_1$ (4-6),

$R_1 + R_2 \le e_1 + e_2$ (4-7),

$2R_1 + R_2 \leq a_1 + g_1 + e_2$ (4-8),

$2R_2 + R_1 \leq a_2 + g_2 + e_1$ (4-9),

$$\left\{\begin{array}{ll} 2R_1 + R_2 \leq 2a_1 + e_2 + f_2 & (4\text{-}10) \\ 2R_2 + R_1 \leq 2a_2 + e_1 + f_1 & (4\text{-}11) \\ R_1 \leq a_1 + e_2 & (4\text{-}12) \\ R_2 \leq a_2 + e_1 & (4\text{-}13) \end{array}\right\}$$

**Proof.** In addition to (4-1,..9) and (4-10),(4-11) in Theorem B [7], in view of $c_i \leq e_i$ , $i = 1,2$ , (4-2) and (4-4) yields (4-12) and (4-13), respectively. Therefore, $(R_1 , R_2)$ satisfies (4-1,…13), where the inequalities (4-10,11,12,13) are redundant.

**The HK region with modified error definition in terms of $(T_1, S_1, T_2, S_2)$**

**Theorem 4** Assuming that the incorrect decoding of $W_1(W_2)$ by the receiver $RX2(RX1)$ is not considered as an error, the region in Theorem 1 is changed as follows, as $S_{IC}^{HK-mod.}(Z)$.

$S_1 \leq I(Y_1; U_1|W_1W_2Q) = a_1$ (3-1),

$T_1 \leq I(Y_1; W_1|U_1W_2Q) = b_1$ (3-2),

~~$T_2 \leq I(Y_1; W_2|U_1W_1Q) = c_1$ (3-3),~~

$S_1 + T_1 \leq I(Y_1; U_1W_1|W_2Q) = d_1$ (3-4),

$S_1 + T_2 \leq I(Y_1; U_1W_2|W_1Q) = e_1$ (3-5),

$T_1 + T_2 \leq I(Y_1; W_1W_2|U_1Q) = f_1$ (3-6),

$S_1 + T_1 + T_2 \leq I(Y_1; U_1W_1W_2|Q) = g_1$ (3-7),

$S_2 \leq I(Y_2; U_2|W_1W_2Q) = a_2$ (3-8),

$T_2 \leq I(Y_2; W_2|U_2W_1Q) = b_2$ (3-9),

~~$T_1 \leq I(Y_2; W_1|U_2W_2Q) = c_2$ (3-10),~~

$S_2 + T_2 \leq I(Y_2; U_2W_2|W_1Q) = d_2$ (3-11),

$S_2 + T_1 \leq I(Y_2; U_2W_1|W_2Q) = e_2$ (3-12),

$T_1 + T_2 \leq I(Y_2; W_1W_2|U_2Q) = f_2$ (3-13),

$S_2 + T_1 + T_2 \leq I(Y_2; U_2W_1W_2|Q) = g_2$ (3-14),

**Proof.** The proof of Theorem 1 is repeated, considering the modified error definition.

**The HK region with modified error definition in terms of $(R_1 , R_2)$**

**Theorem 5** (Theorem C in [7]). Assuming that the incorrect decoding of $W_1(W_2)$ by the receiver $RX2(RX1)$ is not considered as an error, the region in Theorem 4 can be described as follows, as $\mathcal{R}_{HK}^{mod.}$ .

$R_1 \leq d_1$ (8-1)

$R_1 \leq a_1 + e_2$ (8-2)

$R_1 \leq a_1 + f_2$ (8-3)

$R_2 \leq d_2$ (8-4)

$R_2 \leq a_2 + e_1$ (8-5)

$R_2 \leq a_2 + f_1$ (8-6)

$R_1 + R_2 \leq a_1 + g_2$ (8-7)

$R_1 + R_2 \leq a_2 + g_1$ (8-8)

$R_1 + R_2 \leq e_1 + e_2$ (8-9)

$2R_1 + R_2 \leq a_1 + g_1 + e_2$ (8-10)

$2R_1 + R_2 \leq 2a_1 + e_2 + f_2$ (8-11)

$2R_2 + R_1 \leq a_2 + g_2 + e_1$ (8-12)

$2R_2 + R_1 \leq 2a_2 + e_1 + f_1$ (8-13)

**Proof.** We apply the Fourier-Motzkin algorithm to the region in Theorem 4 without the inequalities (3-3,10) in the same manner as in Theorem 2.

Note that in this case, we do not have $c_i$ and hence $c_i \leq e_i$ , $c_i \leq f_i$ , $i = 1,2$. Therefore, (4-2,4) in Theorem 2 are substituted by (8-2,3,5,6) . And also, (8-11,13) are not redundant.

For brevity the details are omitted.

## V. The CMG and modified CMG Rate Regions and the Equivalency with the HK Region

### A. The CMG Region

**The input distribution in the CMG region**

In the CMG coding, the message random variables are independent as in the HK coding. Reduced auxiliary random variables $Q\,, W_1\,, W_2$ are used instead of $Q\,, U_1\,, W_1\,, U_2\,, W_2$, and the message conveyed by $U_1\ (U_2)$ is superimposed over $Q\,, W_1(Q\,, W_2)$ by $X_1(X_2)$. In other words,$(u_1\,, u_2)$ is removed and substituted by $(x_1\,, x_2)$, through the following distribution:

$$p(\,) = p(q)\, p(w_1|q)\, p(x_1|qw_1)\, p(w_2|q)\, p(x_2|qw_2) p(y_1y_2|x_1x_2) \qquad \text{(9-a)},$$

where the input distribution is:

$$p(qw_1w_2x_1x_2) = p(q)\, p(w_1|q)\, p(x_1(u_1)|qw_1)\, p(w_2|q)\, p(x_2(u_2)|qw_2) \qquad \text{(9-b)},$$

where, the notations $x_1(u_1)$ and $x_2(u_2)$ emphasize on superposition of $u_1$and $u_2$ by $x_1$ and $x_2$, respectively.

It is worth noting that, first, due to the independence of $U_i$ and $W_i$ given $Q$, $i = 1,2$ in the HK region, there is always (9-b) for every distribution (2-b):

$$\sum_{u_1u_2} p(qw_1w_2u_1u_2x_1x_2) = p(q)\, p(w_1|q)\, p(x_1(u_1)|qw_1)\, p(w_2|q)\, p(x_2(u_2)|qw_2) \qquad \text{(9-c)}$$

Second, from (9-a) resulting in $(X_1W_1 \longrightarrow Q \longrightarrow W_2X_2\ \ ,\ \ W_1W_2Q \longrightarrow X_1X_2 \longrightarrow Y_1Y_2)$, we have Markov chains:

$$W_1 \longrightarrow QW_2X_1 \longrightarrow Y_1 \qquad \text{(10)},$$

$$W_2 \longrightarrow QW_1X_2 \longrightarrow Y_2 \qquad \text{(11)},$$

Third, obviously and in comparison with the HK coding, the receiver $Y_1$ knowing $W_2X_1$ , knows $W_1$ and the superimposed message $U_1$ and is the case for the receiver $Y_2$ or

$$(U_1), W_1 \longrightarrow QW_2X_1 \longrightarrow Y_1 \qquad \text{(10-a)},$$

$$(U_1), W_1 \longrightarrow QX_1 \longrightarrow Y_1 \qquad \text{(10-b)},$$

$$(U_2), W_2 \longrightarrow QW_1X_2 \longrightarrow Y_2 \qquad \text{(11-a)},$$

$$(U_2), W_2 \longrightarrow QX_2 \longrightarrow Y_2 \qquad \text{(11-b)},$$

**The CMG region in terms of $(T_1, S_1, T_2, S_2)$**

The CMG superposition coding with distribution (9-b), by modifying the error definition and simplifying properties of (10) and (11), leads to the following region which is a simplified version of the HK region.

**Theorem 6** ( Lemma 3, [8] ). For the modified IC in Fig. 2, let $Z_1 = (QW_1W_2X_1X_2Y_1Y_2)$ and let $\mathcal{P}_{IC}^{CMG}$ be the set of all distributions of the form (9-c). For any $Z_1 \in \mathcal{P}_{IC}^{CMG}$ let $S_{IC}^{CMG}(Z_1)$ be the set of all quadruples $(T_1, S_1, T_2, S_2)$ of non-negative real numbers such that

$$S_1 \le I(Y_1;\, X_1|W_1W_2Q) = a'_1 \qquad \text{(12-1)}$$

$$S_1 + T_1 \le I(Y_1;\, X_1|W_2Q) = d'_1 \qquad \text{(12-2)}$$

$$S_1 + T_2 \le I(Y_1;\, X_1W_2|W_1Q) = e'_1 \qquad \text{(12-3)}$$

$$S_1 + T_1 + T_2 \le I(Y_1;\, X_1W_2|Q) = g'_1 \qquad \text{(12-4)}$$

$$S_2 \le I(Y_2;\, X_2|W_2W_1Q) = a'_2 \qquad \text{(12-5)}$$

$$S_2 + T_2 \le I(Y_2;\, X_2|W_1Q) = d'_2 \qquad \text{(12-6)}$$

$$S_2 + T_1 \le I(Y_2;\, X_2W_1|W_2Q) = e'_2 \qquad \text{(12-7)}$$

$$S_2 + T_2 + T_1 \le I(Y_2;\, X_2W_1|Q) = g'_2 \qquad \text{(12-8)},$$

then, any element of the closure of $\bigcup_{Z_1 \in \mathcal{P}_{IC}^{CMG}} S_{IC}^{CMG}(Z_1)$ is achievable.

**Proof.** Refer to [8]. To explain the proof in [8], it is worth noting that the relations (3-3,10) in Theorem 1 are removed due to modifying the error definition. Also, besides (12-1,…,8), we have four relations resulting from (10), (11), in the decoding error analysis of the proof [8]:

$$T_i \le b'_i = d'_i\,,\ \ i = 1,2 \qquad \text{(12-9,10)}$$

$$T_1 + T_2 \le f'_i = g'_i\,, i = 1,2, \qquad \text{(12-11,12)}$$

which become redundant due to (12-2,6) and (12-4,8) respectively. Therefore, for the CMG coding, 14 inequalities in the HK region are reduced to 8 inequalities (12-1,…8).

**The equality of bound constants in the HK and CMG regions**

**Theorem 7.** For $a_i, d_i, e_i, g_i$ , $i = 1,2$ in Theorem 1 and $a'_i$ , $d'_i$ , $e'_i$ , $g'_i$ , $i = 1,2$ in Theorem 6, we have generally:

$$a_1 = I(Y_1;\, U_1|W_1W_2Q) \ge I(Y_1;\, X_1|W_1W_2Q) = a'_1 \qquad \text{(13-1)}$$

$$d_1 = I(Y_1;\, U_1W_1|W_2Q) \ge I(Y_1;\, X_1|W_2Q) = d'_1 = b'_1 \qquad \text{(13-2)}$$

$$e_1 = I(Y_1;\, U_1W_2|W_1Q) \ge I(Y_1;\, X_1W_2|W_1Q) = e'_1 \qquad \text{(13-3)}$$

$$g_1 = I(Y_1;\, U_1W_1W_2|Q) \ge I(Y_1;\, X_1W_2|Q) = g'_1 = f'_1 \qquad \text{(13-4)}$$

$$a_2 = I(Y_2;\, U_2|W_2W_1Q) \ge I(Y_2;\, X_2|W_2W_1Q) = a'_2 \qquad \text{(13-5)}$$

$$d_2 = I(Y_2;\, U_2W_2|W_1Q) \ge I(Y_2;\, X_2|W_1Q) = d'_2 = b'_2 \qquad \text{(13-6)}$$

$$e_2 = I(Y_2;\ U_2W_1|W_2Q) \geq I(Y_2;\ X_2W_1|W_2Q) = e'_2 \qquad (13\text{-}7)$$

$$g_2 = I(Y_2;\ U_2W_2W_1|Q) \geq I(Y_2;\ X_2W_1|Q) = g'_2 = f'_2 \qquad (13\text{-}8),$$

that, equalities (=) hold for the corresponding distributions (2-b) and (9-c).

**Proof.** For the HK coding with encoding function $X_1 = f_1(U_1, W_1|Q)$, we have: $a_1 = I(Y_1;\ U_1|W_1W_2Q) = H(Y_1|W_1W_2Q) - H(Y_1|W_1W_2QU_1) = H(Y_1|W_1W_2Q) - H(Y_1|W_1W_2QU_1X_1) \geq H(Y_1|W_1W_2Q) - H(Y_1|W_1W_2QX_1) = I(Y_1;\ X_1|W_1W_2Q) = a'_1$; and similarly $a_2 \geq a'_2$ and (13-2)-(13-8) are proved.

As explained in [8] (p. 3190, the relation 29), due to the independence of $U_i$ from $W_i$ , $i = 1,2$ in the HK region, there are always corresponding distributions (2-b) , (9-b) with results (10), (10-a) and (11), (11-a) resulting in the (=) in (13-1) - (13-8). Specifically, $X_1 = f_1(U_1, W_1|Q)$ and (10-a) yields $a_1 = a'_1$ and etc.

**The inequalities satisfied by the bound constants in the CMG region**

It is easy to verify, in view of (9-c), (10) and (11) that the following inequalities between the bound constants $a'_i, b'_i = d'_i, e'_i, f'_i = g'_i$ , $i = 1, 2$ in Theorem 6, hold:

$$\begin{cases} a'_i \leq d'_i \leq a'_i + b'_i & , i = 1,2 \\ \quad b'_i = d'_i \end{cases}$$
$$\{a'_i \leq e'_i \qquad , i = 1,2 \qquad \text{(CMG-ineq,eq-1,2,…,20)}$$
$$\{b'_i \leq f'_i \qquad , i = 1,2$$
$$\begin{cases} e'_i, d'_i \leq g'_i \leq b'_i + e'_i\ , a'_i + f'_i & , i = 1,2 \\ \quad f'_i = g'_i \end{cases}$$

**An easily comparable form for the ($T_1, S_1, T_2, S_2$) CMG region**

**Theorem 8.** The CMG region in Theorem 6, in view of the equality of the bound constants in Theorem 7, can be described in the following form comparable to the HK region in Theorem 4:

$$S_1 \leq I(Y_1;\ X_1|W_1W_2Q) = a'_1 = a_1 \qquad (12\text{-}1)$$

~~$T_2 \leq I(Y_1; W_2|X_1Q) = c_1$~~

$$S_1 + T_1 \leq I(Y_1;\ X_1|W_2Q) = d'_1 = d_1 \qquad (12\text{-}2)$$
$$S_1 + T_2 \leq I(Y_1;\ X_1W_2|W_1Q) = e'_1 = e_1 \qquad (12\text{-}3)$$
$$S_1 + T_1 + T_2 \leq I(Y_1;\ X_1W_2|Q) = g'_1 = g_1 \qquad (12\text{-}4)$$
$$S_2 \leq I(Y_2;\ X_2|W_2W_1Q) = a'_2 = a_2 \qquad (12\text{-}5)$$

~~$T_1 \leq I(Y_2; W_1|X_2Q) = c_2$~~

$$S_2 + T_2 \leq I(Y_2;\ X_2|W_1Q) = d'_2 = d_2 \qquad (12\text{-}6)$$
$$S_2 + T_1 \leq I(Y_2;\ X_2W_1|W_2Q) = e'_2 = e_2 \qquad (12\text{-}7)$$
$$S_2 + T_2 + T_1 \leq I(Y_2;\ X_2W_1|Q) = g'_2 = g_2 \qquad (12\text{-}8)$$
$$\left\{\begin{array}{ll} T_1 \leq I(Y_1;\ X_1|W_2Q) = d'_1 = b'_1 = d_1 & (12\text{-}9) \\ T_2 \leq I(Y_2;\ X_2|W_1Q) = d'_2 = b'_2 = d_2 & (12\text{-}10) \\ T_1 + T_2 \leq I(Y_1;\ X_1W_2|Q) = g'_1 = f'_1 = g_1 & (12\text{-}11) \\ T_2 + T_1 \leq I(Y_2;\ X_2W_1|Q) = g'_2 = f'_2 = g_2 & (12\text{-}12) \end{array}\right\}$$

**Proof.** Due to modifying the error definition, the results and the proof of Theorems 6 and 7, the proof is obvious.

**The CMG region in terms of ($R_1$ , $R_2$)**

**Theorem 9** ( Lemma 4,[8] and Theorem D,[7] ). By using the Fourier-Motzkin algorithm, the CMG region in Theorem 6 can be described as $\mathcal{R}_{CMG}$ being the set of $(R_1, R_2)$ satisfying:

$$R_1 \leq d'_1 \qquad (14\text{-}1)$$
$$\boldsymbol{R_1 \leq a'_1 + e'_2} \qquad (14\text{-}2)$$
$$R_2 \leq d'_2 \qquad (14\text{-}3)$$
$$\boldsymbol{R_2 \leq a'_2 + e'_1} \qquad (14\text{-}4)$$
$$R_1 + R_2 \leq a'_1 + g'_2 \qquad (14\text{-}5)$$
$$R_1 + R_2 \leq a'_2 + g'_1 \qquad (14\text{-}6)$$
$$R_1 + R_2 \leq e'_1 + e'_2 \qquad (14\text{-}7)$$
$$2R_1 + R_2 \leq a'_1 + g'_1 + e'_2 \qquad (14\text{-}8)$$
$$2R_2 + R_1 \leq a'_2 + g'_2 + e'_1 \qquad (14\text{-}9),$$

where $a'_i$ , $d'_i$ , $e'_i$ and $g'_i$ , $i = 1,2$ are the same as in Theorem 6.

**Proof.** Refer to [7].

**An easily comparable form for the CMG $(\boldsymbol{R_1}, \boldsymbol{R_2})$ region**

Note that in both Theorems 4,8, we do not have $c_i$ , $i = 1,2$, and hence, in a mathematically comparison, their $(R_1 ,\ R_2)$ versions (Theorems 5,9) must be the same. However, due to the differences on the corresponding inequalities related to the bound constants, $(R_1 ,\ R_2)$ regions are different as below.

**Theorem 10.** Explanatory and easily comparable form of $\mathcal{R}_{CMG}$ in Theorem 9 can be described as $\mathcal{R}_{CMG-equi}$. (comparable to $\mathcal{R}_{HK}^{mod}$. in Theorem 5) satisfying thirteen relations for $(R_1 ,\ R_2)$.

$$R_1 \leq d'_1 \quad (14\text{-}1)$$
$$\boldsymbol{R_1 \leq a'_1 + e'_2} \quad (14\text{-}2)$$
$$R_2 \leq d'_2 \quad (14\text{-}3)$$
$$\boldsymbol{R_2 \leq a'_2 + e'_1} \quad (14\text{-}4)$$
$$R_1 + R_2 \leq a'_1 + g'_2 \quad (14\text{-}5)$$
$$R_1 + R_2 \leq a'_2 + g'_1 \quad (14\text{-}6)$$
$$R_1 + R_2 \leq e'_1 + e'_2 \quad (14\text{-}7)$$
$$2R_1 + R_2 \leq a'_1 + g'_1 + e'_2 \quad (14\text{-}8)$$
$$2R_2 + R_1 \leq a'_2 + g'_2 + e'_1 \quad (14\text{-}9)$$
$$\left\{\begin{array}{ll} R_1 \leq a'_1 + f'_2 & (14\text{-}10) \\ R_2 \leq a'_2 + f'_1 & (14\text{-}11) \\ 2R_1 + R_2 \leq 2a'_1 + e'_2 + f'_2 & (14\text{-}12) \\ 2R_2 + R_1 \leq 2a'_2 + e'_1 + f'_1 & (14\text{-}13) \end{array}\right\}$$

**Proof.** According to (CMG-ineq,equ-1,2,…,20), it is obvious that (14-10), (14-11) are redundant due to (14-2),(14-4), respectively. And also, (14-2,5) and (14-4,6) yield (14-12) and (14-13), respectively.

**B. Modified CMG Region**

The CMG region, at first, was claimed to be a new region due to its seemingly difference with the HK region, as you see in Theorems (1,6) and (2,9). Later, it was proved that the two regions are equivalent.

Now, we define the CMG region without modifying the error definition as the modified CMG region the equivalency of which with the HK region is readily seen.(See Theorems 2,12 instead of Theorems 2,9.) Aiming at this definition, we have the following theorems for this region.

**Theorem 11.** Modified CMG region $S_{IC}^{mod-CMG}$ is the set of all quadruples $(T_1, S_1, T_2, S_2)$ of non-negative real numbers such that

$$S_1 \leq I(Y_1;\ X_1|W_1W_2Q) = a'_1 \quad (12\text{-}1)$$
$$T_2 \leq I(Y_1; W_2|X_1Q) = c'_1 \quad \textbf{(12-a)}$$
$$S_1 + T_1 \leq I(Y_1;\ X_1|W_2Q) = d'_1 \quad (12\text{-}2)$$
$$S_1 + T_2 \leq I(Y_1;\ X_1W_2|W_1Q) = e'_1 \quad (12\text{-}3)$$
$$S_1 + T_1 + T_2 \leq I(Y_1;\ X_1W_2|Q) = g'_1 \quad (12\text{-}4)$$
$$S_2 \leq I(Y_2;\ X_2|W_2W_1Q) = a'_2 \quad (12\text{-}5)$$
$$T_1 \leq I(Y_2; W_1|X_2Q) = c'_2 \quad \textbf{(12-b)}$$
$$S_2 + T_2 \leq I(Y_2;\ X_2|W_1Q) = d'_2 \quad (12\text{-}6)$$
$$S_2 + T_1 \leq I(Y_2;\ X_2W_1|W_2Q) = e'_2 \quad (12\text{-}7)$$
$$S_2 + T_2 + T_1 \leq I(Y_2;\ X_2W_1|Q) = g'_2 \quad (12\text{-}8)$$

**Proof.** The proof is the same as the proof of Theorem 6. It is worth noting that, here, we have ten inequalities rather than eight ones in Theorem 6. Also, as in Theorem 6, modified CMG region like the CMG region has redundant inequalities (12-9,10,11,12).

**The inequalities satisfied by the bound constants in the modified CMG region**

It is easy to verify, in view of (9-c), (10) and (11) that the following inequalities between the bound constants $a'_i, b'_i = d'_i, c'_i, e'_i, f'_i = g'_i$ , $i = 1, 2$ in Theorem 6 and or Theorem 11 hold:

$$\begin{array}{ll} \begin{cases} a'_i \leq d'_i \leq a'_i + b'_i \\ \quad b'_i = d'_i \end{cases} & ,\ i = 1,2 \\ \begin{cases} a'_i \leq e'_i \leq a'_i + c'_i \\ \quad c'_i \leq e'_i \end{cases} & ,\ i = 1,2 \\ \{ b'_i \leq f'_i \leq b'_i + c'_i & ,\ i = 1,2 \\ \begin{cases} e'_i, d'_i \leq g'_i \leq b'_i + e'_i\ , a'_i + f'_i \\ \quad f'_i = g'_i \end{cases} & ,\ i = 1,2 \end{array} \quad \text{(mod-CMG-ineq,eq-1,2,…,26)}$$

**The mod-CMG region in terms of ($R_1$ , $R_2$)**

Among the above inequalities, the inequality $c'_i \le e'_i$ and the result $c'_i + g'_i \le e'_i + f'_i$ , determine the mod-CMG region as follows.

**Theorem 12.** The region in Theorem 11, by using the Fouriet-Motzkin elimination technique, can be described as $\mathcal{R}_{mod-CMG}$ being the set of $(R_1, R_2)$ satisfying:

$$R_1 \le d'_1 \quad (4\text{-}1\text{-m}),$$
$$\boldsymbol{R_1 \le a'_1 + c'_2} \quad (4\text{-}2\text{-m}),$$
$$R_2 \le d'_2 \quad (4\text{-}3\text{-m}),$$
$$\boldsymbol{R_2 \le a'_2 + c'_1} \quad (4\text{-}4\text{-m}),$$
$$R_1 + R_2 \le a'_1 + g'_2 \quad (4\text{-}5\text{-m}),$$
$$R_1 + R_2 \le a'_2 + g'_1 \quad (4\text{-}6\text{-m}),$$
$$R_1 + R_2 \le e'_1 + e'_2 \quad (4\text{-}7\text{-m}),$$
$$2R_1 + R_2 \le a'_1 + g'_1 + e'_2 \quad (4\text{-}8\text{-m}),$$
$$2R_2 + R_1 \le a'_2 + g'_2 + e'_1 \quad (4\text{-}9\text{-m}),$$

where $a'_i, c'_i, d'_i$, $e'_i$ and $g'_i$ , $i = 1,2$ are the same as in Theorem 11.

**Proof.** In view of (mod-CMG-ineq,equ-1,2,…,26), the proof is identical to the proof of Theorem 2.

**C. Equivalency of the HK, the CMG and mod-CMG regions**

The regions can be compared with the help of the above twelve theorems in detail. It can be proved that the HK, the CMG and mod-CMG regions are equivalent.

**C.1** Equivalency of the HK and mod- CMG regions

By considering Theorems 2 and 12, due to the equalities in Theorem 7 and the equality of $c'_i$ in Theorem 11 with $c_i$ in Theorem 1, $= 1,2$ , the HK and mod-CMG regions are easily proved to be equivalent, in contrast to the dissimilarity in Theorems (2,9) resulting in the hardness to prove the equivalency.

**C.2** Equivalency of the HK and the CMG regions

As mentioned above, to prove the equivalency of the HK and the CMG regions is hard as you see below.

**Theorem 13.**. The HK and the CMG $(R_1, R_2)$ regions are equivalent such that:

$(R_1, R_2) \in \mathcal{R}_{HK} =$ (4-1,…,9) or (4-1,…,13) $\Leftrightarrow (R_1, R_2) \in \mathcal{R}_{CMG} =$ (14-1,…,9) $\Leftrightarrow (R_1, R_2) \in$ (15-1,…,7), where (15-1,…,7) are as follows.

$$R_1 \le d'_1 = d_1 \quad (15\text{-}1)$$
$$R_2 \le d'_2 = d_2 \quad (15\text{-}2)$$
$$R_1 + R_2 \le a'_1 + g'_2 = a_1 + g_2 \quad (15\text{-}3)$$
$$R_1 + R_2 \le a'_2 + g'_1 = a_2 + g_1 \quad (15\text{-}4)$$
$$R_1 + R_2 \le e'_1 + e'_2 = e_1 + e_2 \quad (15\text{-}5)$$
$$2R_1 + R_2 \le a'_1 + g'_1 + e'_2 = a_1 + g_1 + e_2 \quad (15\text{-}6)$$
$$2R_2 + R_1 \le a'_2 + g'_2 + e'_1 = a_2 + g_2 + e_1 \quad (15\text{-}7).$$

**Proof.** Due to the independence of $U_i$ from $W_i$ , $i = 1,2$ in the HK region, there are always corresponding distributions (2-b) (with $X_i = f_i(U_i, W_i|Q)$ , $i = 1,2$) and (9-b) (with results (10), (11)), thereby holding (=) in Theorem 7 and resulting in the above equivalences, as proved in [8] ( Theorem 2 for (15-1)-(15-7) ); Lemma 1 for (4-1)-(4-9); Lemma 4 for (14-1)-(14-9); Lemma 2 for proving the equivalency of the two regions).

In [8], it is proved that the rate region (15-1,..7) is the union of three HK regions (the HK region, the HK region with $W_1 = \emptyset$, the HK region with $W_2 = \emptyset$, ) for three different superposition coding strategies. Also, it can be proved that the region (15-1,..7) is obtained as the union of three rate regions in Theorem 6 ($S_{IC}^{CMG}$, $S_{IC}^{CMG}(\ W_1 = \emptyset), S_{IC}^{CMG}(\ W_2 = \emptyset)$ for three different superposition coding schemes [28].

## VI. NEW RATE REGION FOR GENERAL INTERFERENCE CHANNEL ( The Hodtani Region)

As mentioned in section I, we make two novel changes in the HK and CMG coding. First, we allow input auxiliary random variables to be correlated and second, exploit random binning technique and jointly decoding strategy, in order to obtain a new rate region for general IC and to improve the HK and CMG regions, where message variables are independent. We describe our work in the following parts.

(i) To explain the input distribution.
(ii) To obtain the new rate region in terms of $(T_1, S_1, T_2, S_2)$ .
(iii) To interpret the region and to intuitively see its newness
(iv) To determine the inequalities satisfied by the bound constants in our region in contrast to polymatroidal inequalities of the HK region.
(v) To describe the region in terms of $(R_1, R_2)$.

(vi) To obtain our $(R_1, R_2)$ region with modified error definition.

**(i) The input distribution in the Hodtani rate region**

We consider the general distribution (1) for the IC or allow the auxiliary variables in the HK distribution (2) to be correlated in the following form:

$$p(\,) = p(q)\, p(w_{1d}|q)\, p(u_{1d}|qw_{1d})\, p(w_{2d}|q)\, p(u_{2d}|qw_{2d}) p_q(y_1y_2|u_{1d}w_{1d}u_{2d}w_{d2}) \qquad (16),$$

where $p_q(y_1y_2|u_{1d}w_{1d}u_{2d}w_{2d}) = p\big(y_1y_2\big|x_1 = f_1(u_{1d}w_{1d}|q), x_2 = f_2(u_{2d}w_{2d}|q)\big)$, and $d$ in the variables $u_{1d}w_{1d}u_{2d}w_{2d}$ denotes the dependency, i.e., $(u_{1d}, w_{1d})$ , $(u_{2d}, w_{2d})$ are dependent random variables, in contrast to the HK-CMG distributions with independent message variables in (2), (9).

**(ii) Hodtani rate region for the IC**

Now, we obtain our achievable rate region for (16) as follows

**Theorem 14.** For the modified IC (Fig.2), let $Z_2 = (QU_{1d}W_{1d}U_{2d}W_{2d}X_1X_2Y_1Y_2)$ and let $\mathcal{P}_{IC}^{Hod}$ be the set of all distributions of the form (16). For any $Z_2 \in \mathcal{P}_{IC}^{Hod}$ let $S_{IC}^{Hod}(Z_2)$ be the set of all quadruples $(T_1, S_1, T_2, S_2)$ of nonnegative real numbers such that

$$S_1 \leq I(Y_1; U_{1d}|W_{1d}W_{2d}Q) = A_1 \qquad (17\text{-}1),$$
$$T_1 \leq I(Y_1; W_{1d}|U_{1d}W_{2d}Q) + I(U_{1d}; W_{1d}|Q) = B_1 \qquad (17\text{-}2),$$
$$T_2 \leq I(Y_1; W_{2d}|U_{1d}W_{1d}Q) + I(U_{1d}; W_{1d}|Q) = C_1 \qquad (17\text{-}3),$$
$$S_1 + T_1 \leq I(Y_1; U_{1d}W_{1d}|W_{2d}Q) = D_1 \qquad (17\text{-}4),$$
$$S_1 + T_2 \leq I(Y_1; U_{1d}W_{2d}|W_{1d}Q) = E_1 \qquad (17\text{-}5),$$
$$T_1 + T_2 \leq I(Y_1; W_{1d}W_{2d}|U_{1d}Q) + I(U_{1d}; W_{1d}|Q) = F_1 \qquad (17\text{-}6),$$
$$S_1 + T_1 + T_2 \leq I(Y_1; U_{1d}W_{1d}W_{2d}|Q) = G_1 \qquad (17\text{-}7),$$
$$S_2 \leq I(Y_2; U_{2d}|W_{1d}W_{2d}Q) = A_2 \qquad (17\text{-}8),$$
$$T_2 \leq I(Y_2; W_{2d}|U_{2d}W_{1d}Q) + I(U_{2d}; W_{2d}|Q) = B_2 \qquad (17\text{-}9),$$
$$T_1 \leq I(Y_2; W_{1d}|U_{2d}W_{2d}Q) + I(U_{2d}; W_{2d}|Q) = C_2 \qquad (17\text{-}10),$$
$$S_2 + T_2 \leq I(Y_2; U_{2d}W_{2d}|W_{1d}Q) = D_2 \qquad (17\text{-}11),$$
$$S_2 + T_1 \leq I(Y_2; U_{2d}W_{1d}|W_{2d}Q) = E_2 \qquad (17\text{-}12),$$
$$T_1 + T_2 \leq I(Y_2; W_{1d}W_{2d}|U_{2d}Q) + I(U_{2d}; W_{2d}|Q) = F_2 \qquad (17\text{-}13),$$
$$S_2 + T_1 + T_2 \leq I(Y_2; U_{2d}W_{1d}W_{2d}|Q) = G_2 \qquad (17\text{-}14),$$

then, any element of the closure of $\bigcup_{Z_2 \in \mathcal{P}_{IC}^{Hod}} S_{IC}^{Hod}(Z_2)$ is achievable, where $A_i, B_i, C_i, D_i, E_i, F_i, G_i$ , $i = 1{,}2$ all differ from the corresponding terms $a_i, b_i, c_i, d_i, e_i, f_i, g_i$ , $i = 1{,}2$ in the HK-CMG region (see the Theorem 22 in section VIII).

**Proof:** Refer to Appendix A.

**(iii) An interesting interpretation of the Hodtani region (to intuitively explain the differences between the HK and the Hodtani regions) :**

In accordance with the distribution (16), we have allowed $u_{1d}w_{1d}$ (and also $u_{2d}w_{2d}$) to be correlated and used a binning scheme (see the proof of Theorem 14 in Appendix). Therefore, we have added two additional terms to every rate in the HK region, with dependent variables: One positive term $I(U_{1d}; W_{1d}|Q)$ or $I(U_{2d}; W_{2d}|Q)$ indicating the input correlation, and one negative term $-I(U_{1d}; W_{1d}|Q)$ or $-I(U_{2d}; W_{2d}|Q)$ illustrating the binning scheme.

In the rates including $S_1$ or $S_2$, where we have applied the binning scheme, we have both positive and negative terms cancelling each other. In the rates including $T_1$ and or $T_2$, where we don not have the binning scheme, we see only additional positive term $I(U_{1d}; W_{1d}|Q)$ or $I(U_{2d}; W_{2d}|Q)$ illustrating the dependency of the variables in our coding.

As it will be proved in Theorem 22, based on the the Wyner common information, that all of the rate terms in our region are greater than the corresponding terms in the HK-CMG region.

**(iv) The inequalities satisfied by the bound constants in the Hodtani region**

We prove that six inequalities of 30 polymatroidal inequalities (HK-ineq- 1,2,…,30) of the HK-CMG bound constants in section IV are not satisfied by the bound constants in our region as follows.

**Theorem 15.** For the bound constants $A_i, B_i, C_i, D_i, E_i, F_i, G_i$ , $i = 1{,}2$ in Theorem 14, we have:

$$\begin{cases} A_i \leq D_i \leq A_i + B_i \quad , \; i = 1{,}2 \\ \cancel{B_i \leq D_i} \end{cases}$$

$$\begin{cases} A_i \leq E_i \leq A_i + C_i \quad , \; i = 1{,}2 \\ \cancel{C_i \leq E_i} \end{cases} \qquad \text{(Hod-ineq 1,…,24)}$$

$$\{C_i\,, B_i \le F_i \le B_i + C_i \qquad , \; i = 1,2$$

$$\begin{cases} E_i\,, D_i \le G_i \le C_i + D_i\,, B_i + E_i\,, A_i + F_i \qquad , \; i = 1,2 \\ \xcancel{F_i \le G_i} \end{cases}$$

**Proof.** It is sufficient to prove the theorem for $i = 1$.

- $A_1 \le D_1 \le A_1 + B_1$

From theorem 14, we have:

$D_1 = I(Y_1;\, U_{1d}W_{1d}|W_{2d}Q) = I(Y_1; W_{1d}|W_{2d}U_{1d}Q) + I(Y_1; U_{1d}|QW_{2d}) =$

$I(Y_1; W_{1d}|QW_{2d}U_{1d}) + H(U_{1d}|QW_{2d}) - H(U_{1d}|QW_{2d}Y_1) \overset{a}{\le}$

$I(Y_1; W_{1d}|QW_{2d}U_{1d}) + H(U_{1d}|Q) - H(U_{1d}|QW_{2d}Y_1W_{1d}) \overset{b}{=}$

$I(Y_1; W_{1d}|QW_{2d}U_{1d}) + H(U_{1d}|Q) - H(U_{1d}|QW_{2d}Y_1W_{1d}) + H(U_{1d}|W_{1d}W_{2d}Q) - H(U_{1d}|QW_{1d}) =$

$I(Y_1; W_{1d}|QW_{2d}U_{1d}) + H(U_{1d}|Q) - H(U_{1d}|QW_{1d}) + H(U_{1d}|QW_{1d}W_{2d}) - H(U_{1d}|QW_{2d}W_{1d}Y_1) \overset{c}{=}$

$I(U_{1d}; W_{1d}|Q) + I(Y_1; W_{1d}|QW_{2d}U_{1d}) + I(Y_1; U_{1d}|QW_{2d}W_{1d}) = B_1 + A_1$ ,

where, a,b,c follows from $H(X|YZ) \le H(X|Y)$ , $U_{1d} \longrightarrow QW_{1d} \longrightarrow W_{2d}$ and $I(X;Y) = H(X) - H(X|Y)$, respectively. And, $D_1 \ge A_1$ is obvious.

In order to verify $D_1 \ge B_1$ , the inequality $I(Y_1; U_{1d}|QW_{2d}) \ge I(U_{1d}; W_{1d}|Q)$ is necessary that is not satisfied for dependent $(U_{1d}, W_{1d})$ . Therefore, ~~$B_1 \le D_1$~~ .

- $E_1 \le A_1 + C_1$

From theorem 14, we have:

$E_1 = I(Y_1;\, U_{1d}W_{2d}|QW_{1d}) = I(Y_1;\, U_{1d}|QW_{1d}) + I(Y_1; W_{2d}|QW_{1d}U_{1d})$, and

$A_1 + C_1 = I(Y_1; U_{1d}|QW_{1d}W_{2d}) + I(Y_1; W_{2d}|QW_{1d}U_{1d}) + I(U_{1d}; w_{1d}|Q)$

Due to $U_{1d} \longrightarrow QW_{1d} \longrightarrow W_{2d}$ , $H(X|YZ) \le H(X|Y)$ and $I(X;Y) \ge 0$ , we have:

$I(Y_1; U_{1d}|QW_{1d}) = H(U_{1d}|QW_{1d}) - H(U_{1d}|QW_{1d}Y_1) \le$

$H(U_{1d}|QW_{1d}W_{2d}) - H(U_{1d}|QW_{1d}W_{2d}Y_1) + I(U_{1d}; W_{1d}|Q) \le I(Y_1; U_{1d}|QW_{1d}W_{2d}) + I(U_{1d}; W_{1d}|Q)$ ,

then, $E_1 \le A_1 + C_1$ holds. Similarly, $E_1 \ge A_1$ obviously holds; however, the inequality $E_1 \ge C_1$ does not hold for dependent $(U_{1d}, W_{1d})$.

- The other relations in the theorem hold and the verification does not have anything new and the details are omitted.

**(v) The Hodtani region in terms of ($\boldsymbol{R_1}$ , $\boldsymbol{R_2}$)**

Now, we describe our region as $(R_1\,, R_2)$ rates.

**Theorem 16.** The $S_{IC}^{Hod}(Z_2)$ region in Theorem 14 can be described, using the Fourier-Motzkin algorithm, as $\mathcal{R}_{Hod}$ being the set of $(R_1\,, R_2)$ satisfying:

$$R_1 \le D_1 \qquad (18\text{-}1)$$

$$R_1 \le A_1 + C_2 \qquad (18\text{-}2)$$

$$\boldsymbol{R_1 \le A_1 + E_2} \qquad (18\text{-}3)$$

$$R_2 \le D_2 \qquad (18\text{-}4)$$

$$R_2 \le A_2 + C_1 \qquad (18\text{-}5)$$

$$\boldsymbol{R_2 \le A_2 + E_1} \qquad (18\text{-}6)$$

$$R_1 + R_2 \le A_2 + G_1 \qquad (18\text{-}7)$$

$$R_1 + R_2 \le A_1 + G_2 \qquad (18\text{-}8)$$

$$R_1 + R_2 \le E_1 + E_2 \qquad (18\text{-}9)$$

$$2R_1 + R_2 \le A_1 + G_1 + E_2 \qquad (18\text{-}10)$$

$$\boldsymbol{2R_1 + R_2 \le 2A_1 + E_2 + F_2} \qquad (18\text{-}11)$$

$$2R_2 + R_1 \leq A_2 + G_2 + E_1 \quad (18\text{-}12)$$
$$\boldsymbol{2R_2 + R_1 \leq 2A_2 + E_1 + F_1} \quad (18\text{-}13),$$

where, the bound constants $A_i, B_i, C_i, D_i, E_i, F_i, G_i$ , $i = 1,2$ are the same as in Theorem 14.

**Proof.** Refer to Appendix B.

**Outline of the proof**.This theorem is proved by virtue of the Fourier-Motzkin elimination technique in the same manner as in Theorem 2 (the details are in Appendix B), but with two differences:

First, here, the inequalities $C_i \leq E_i$ , $i = 1,2$ are not satisfied, and hence there remain $R_1 \leq A_1 + E_2$ , $R_2 \leq A_2 + E_1$, in addition to the inequalities in Theorem 2.

Second, in our general distribution (16), $U_i, W_i$ , $i = 1,2$ are not independent given $Q$, the result of which is violating (7). Therefore, the inequalities (18-11) and (18-13) remain and are not redundant due to (18-2,7) and (18-5,8), respectively, despite the case in Theorem 2 for the HK region.

**(vi) The Hodtani $(\boldsymbol{R_1}, \boldsymbol{R_2})$ region with modified error definition**

As we have described the HK region with modified error definition in terms of $(T_1, S_1, T_2, S_2)$ in Theorem 4, $(R_1, R_2)$ in Theorem 5, now we obtain our region only in $(R_1, R_2)$ with modifying the error definition.

**Theorem 17.** Assuming that the incorrect decoding of $W_1(W_2)$ by the receiver $RX2(RX1)$ is not considered as an error, the region in Theorem 14 can be described as follows, as $\mathcal{R}_{Hod}^{mod\cdot}$ .

$$R_1 \leq D_1 \quad (19\text{-}1)$$
$$R_1 \leq A_1 + E_2 \quad (19\text{-}2)$$
$$R_1 \leq A_1 + F_2 \quad (19\text{-}3)$$
$$R_2 \leq D_2 \quad (19\text{-}4)$$
$$R_2 \leq A_2 + E_1 \quad (19\text{-}5)$$
$$R_2 \leq A_2 + F_1 \quad (19\text{-}6)$$
$$R_1 + R_2 \leq A_1 + G_2 \quad (19\text{-}7)$$
$$R_1 + R_2 \leq A_2 + G_1 \quad (19\text{-}8)$$
$$R_1 + R_2 \leq E_1 + E_2 \quad (19\text{-}9)$$
$$2R_1 + R_2 \leq 2A_1 + F_2 + E_2 \quad (19\text{-}10)$$
$$2R_1 + R_2 \leq G_1 + A_1 + E_2 \quad (19\text{-}11)$$
$$2R_2 + R_1 \leq 2A_2 + E_1 + F_1 \quad (19\text{-}12)$$
$$2R_2 + R_1 \leq A_2 + E_1 + G_2 \quad (19\text{-}13)$$

where, the bound constants $A_i, B_i, C_i, D_i, E_i, F_i, G_i$ , $i = 1,2$ are the same as in Theorem 14.

**Proof.** By considering the rates in Theorem 14 without (17-3,10) and applying Fourier-Motzkin elimination technique the same as in the proof for Theorem 16 in Appendix B, and then removing redundant inequalities in view of Theorem 15, we reach readily to (19-1,…13). As expected, (19-1,…13) are (8-1,…,13), where $a_i, b_i, c_i, d_i, e_i, f_i$ and $g_i$ , $i = 1,2$ in the HK region are replaced in our region by $A_i, B_i, C_i, D_i, E_i, F_i, G_i$ , $i = 1,2$, respectively.

## VII. The HK-CMG Region and The Hodtani-CMG Region

For comparison purposes and in order to exactly demarcate the regions, first, we redefine the CMG region as the HK-CMG region. Second, we define the Hodtani-CMG region as a new and improved CMG region and obtain different versions for this region.

### a. The HK-CMG region

**Definition.**The HK-CMG region is defined to be the CMG region in section V obtained through distribution (9-c) which is the result of adding the HK distribution (2-b) with independent $(w_1, w_2, u_1, u_2)$ over $(u_1, u_2)$ as in (9-c).

The input distribution in the HK-CMG region is (9-c), where the message variables are independent. Therefore, the HK-CMG region is obtained using superposition coding-jointly decoding and is the same as the CMG region and hence, is equivalent to the HK region as stated before in Theorem 13.

Now, we make the message variables dependent in the CMG region and define a new region as the Hodtani-CMG region which is an improved version for the CMG region.

### b. The Hodtani-CMG region

**Definition**. The Hodtani-CMG region is defined to be the CMG region obtained through the Hodtani input distribution (16) as follows.

$$\sum_{u_{1d}u_{2d}} p(qw_{1d}w_{2d}u_{1d}u_{2d}x_1x_2) = p(qw_{1d}w_{2d}x_1x_2) = p(q)\,p(w_{1d}|q)\,p(x_1|w_{1d}q)\ p(w_{2d}|q)\,p(x_2|w_{2d}q) \quad (20),$$

where, the message variables $(w_{id}, u_{id})$, $i = 1,2$ are dependent, whereas $(w_i, u_i)$, $i = 1,2$ in (9-c) are independent.

The noticeable contrast between the Hodtani-CMG region and the HK and the HK-CMG regions is this dependency. In order to explain (20), we derive its different versions.

The other versions of (20):
In applying superposition coding, the message superimposed over $w_{id}$ , $i = 1,2$ by $x_i$ , $i = 1,2$ is independent of $w_{id}$ or in our notation ($w_{id} = (w_i, k_i)$ , $u_{id} = (u_i, k_i)$ , $i = 1,2$ , with independent $w_i, u_i, k_i$ , $i = 1,2$):

$$\sum_{u_1 u_2} p(q w_1 k_1 w_2 k_2 u_1 k_1 u_2 k_2 x_1 x_2) = p(q w_{1d} w_{2d} x_1 x_2)$$
$$= p(q)\, p(w_{1d}|q)\, p(x_1(u_1)|w_{1d} q)\ \ p(w_{2d}|q)\, p(x_2(u_2)|w_{2d} q) \qquad (21),$$

and also,the relation (21) can be written as follows.
.

$$\sum_{u_1 u_2 k_1 k_2} p(q w_1 k_1 w_2 k_2 u_1 k_1 u_2 k_2 x_1 x_2) = p(q w_1 w_2 x_1 x_2)$$
$$= p(q)\, p(w_1|q)\, p(x_1(u_1 k_1)|w_1 q)\ \ p(w_2|q)\, p(x_2(u_2 k_2)|w_2 q) \qquad (22),$$

and, mathematically, in general,

$$\textstyle\sum_{u_{1d} u_{2d}} p(q w_{1d} w_{2d} u_{1d} u_{2d} x_1 x_2) = p(q w_{1d} w_{2d} x_1 x_2) =$$
$$p(q)\, p(w_{1d}|q)\, p(x_1(u_{1d})|w_{1d} q)\ \ p(w_{2d}|q)\, p(x_2(u_{2d})|w_{2d} q) \qquad (23)$$

As for (9-c), from (23) resulting in ($X_1 W_{1d} \longrightarrow Q \longrightarrow W_{2d} X_2$ , $W_{1d} W_{2d} Q \longrightarrow X_1 X_2 \longrightarrow Y_1 Y_2$), we have Markov chains:

$$W_{1d} \longrightarrow Q W_{2d} X_1 \longrightarrow Y_1 \qquad (24),$$
$$W_{2d} \longrightarrow Q W_{1d} X_2 \longrightarrow Y_2 \qquad (25).$$

Obviously and in comparison with the Hodtani coding, the receiver $Y_1$ knowing $W_{2d} X_1$ , knows $W_{1d}$ and the superimposed message $U_{1d}$ and is the case for the receiver $Y_2$ or

$$(U_{1d}), W_{1d} \longrightarrow Q W_{2d} X_1 \longrightarrow Y_1 \qquad (24\text{-a}),$$
$$(U_{1d}), W_{1d} \longrightarrow Q X_1 \longrightarrow Y_1 \qquad (24\text{-b}),$$
$$(U_{2d}), W_{2d} \longrightarrow Q W_{1d} X_2 \longrightarrow Y_2 \qquad (25\text{-a}),$$
$$(U_{2d}), W_{2d} \longrightarrow Q X_2 \longrightarrow Y_2 \qquad (25\text{-b}),$$

**The Hodtani-CMG region in terms of ($T_1, S_1, T_2, S_2$)**

According to (20) and or its other versions, by using superposition coding, the new region is obtained as follows.

**Theorem 18.** For the modified IC in Fig. 2, let $Z_3 = (Q W_{1d} W_{2d} X_1 X_2 Y_1 Y_2)$ and let $\mathcal{P}_{IC}^{Hod-CMG}$ be the set of all distributions of the form (23). For any $Z_3 \in \mathcal{P}_{IC}^{Hod-CMG}$ let $S_{IC}^{Hod-CMG}(Z_3)$ be the set of all quadruples $(T_1, S_1, T_2, S_2)$ of non-negative real numbers such that

$$S_1 \le I(Y_1;\, X_1|W_{1d} W_{2d} Q) = A'_1 \qquad (26\text{-}1)$$
$$S_1 + T_1 \le I(Y_1;\, X_1|W_{2d} Q) = D'_1 \qquad (26\text{-}2)$$
$$S_1 + T_2 \le I(Y_1;\, X_1 W_{2d}|W_{1d} Q) = E'_1 \qquad (26\text{-}3)$$
$$S_1 + T_1 + T_2 \le I(Y_1;\, X_1 W_{2d}|Q) = G'_1 \qquad (26\text{-}4)$$
$$S_2 \le I(Y_2;\, X_2|W_{2d} W_{1d} Q) = A'_2 \qquad (26\text{-}5)$$
$$S_2 + T_2 \le I(Y_2;\, X_2|W_{1d} Q) = D'_2 \qquad (26\text{-}6)$$
$$S_2 + T_1 \le I(Y_2;\, X_2 W_{1d}|W_{2d} Q) = E'_2 \qquad (26\text{-}7)$$
$$S_2 + T_2 + T_1 \le I(Y_2;\, X_2 W_{1d}|Q) = G'_2 \qquad (26\text{-}8),$$

then, any element of the closure of $\bigcup_{Z_3 \in \mathcal{P}_{IC}^{Hod-CMG}} S_{IC}^{Hod-CMG}(Z_3)$ is achievable.

**Proof.** By modifying the error definition and simplifying properties of (24) and (25), the region is obtained in the same manner as in Theorem 6. The details are omitted.

**The equality of bound constants in the Hodtani region and the Hodtani- CMG region**

**Theorem 19.** For $A_i, D_i, E_i, G_i$ , $i = 1,2$ in Theorem 14 and $A'_i$ , $D'_i$ , $E'_i$ , $G'_i$ , $i = 1,2$ in Theorem 18, we have:

$$A_1 = I(Y_1;\, U_{1d}|W_{1d} W_{2d} Q) = I(Y_1;\, X_1|W_{1d} W_{2d} Q) = A'_1 \qquad (27\text{-}1)$$
$$D_1 = I(Y_1;\, U_{1d} W_{1d}|W_{2d} Q) = I(Y_1;\, X_1|W_{2d} Q) = D'_1 = B'_1 \qquad (27\text{-}2)$$
$$E_1 = I(Y_1;\, U_1 W_{2d}|W_{1d} Q) = I(Y_1;\, X_1 W_{2d}|W_{d1} Q) = E'_1 \qquad (27\text{-}3)$$
$$G_1 = I(Y_1;\, U_{1d} W_{d1} W_{2d}|Q) = I(Y_1;\, X_1 W_{2d}|Q) = G'_1 = F'_1 \qquad (27\text{-}4)$$
$$A_2 = I(Y_2;\, U_2|W_{2d} W_{1d} Q) = I(Y_2;\, X_2|W_{2d} W_{1d} Q) = A'_2 \qquad (27\text{-}5)$$
$$D_2 = I(Y_2;\, U_{2d} W_{2d}|W_{1d} Q) = I(Y_2;\, X_2|W_{1d} Q) = D'_2 = B'_2 \qquad (27\text{-}6)$$
$$E_2 = I(Y_2;\, U_{2d} W_{1d}|W_{2d} Q) = I(Y_2;\, X_2 W_{1d}|W_{2d} Q) = E'_2 \qquad (27\text{-}7)$$
$$G_2 = I(Y_2;\, U_{2d} W_{2d} W_{1d}|Q) = I(Y_2;\, X_2 W_{1d}|Q) = G'_2 = F'_2 \qquad (27\text{-}8),$$

**Proof.** The equalities in Theorem 19 and Theorem 7 are similar, therefore, in view of the mathematical similarity between (9-c) and (23), and also, between (10-a, 11-a) and (24-a, 25-a), the proof is the same as the proof for Theorem 7.

**The Hodtani- CMG region in terms of ($\boldsymbol{R_1}$ , $\boldsymbol{R_2}$)**

**Theorem 20.** By using the Fourier-Motzkin algorithm, the Hodtani-CMG region in Theorem 18 can be described as $\mathcal{R}_{H0d-CMG}$ being the set of $(R_1 , R_2)$ satisfying:

$$R_1 \leq D'_1 \quad (28\text{-}1)$$
$$\boldsymbol{R_1 \leq A'_1 + E'_2} \quad (28\text{-}2)$$
$$R_2 \leq D'_2 \quad (28\text{-}3)$$
$$\boldsymbol{R_2 \leq A'_2 + E'_1} \quad (28\text{-}4)$$
$$R_1 + R_2 \leq A'_1 + G'_2 \quad (28\text{-}5)$$
$$R_1 + R_2 \leq A'_2 + G'_1 \quad (28\text{-}6)$$
$$R_1 + R_2 \leq E'_1 + E'_2 \quad (28\text{-}7)$$
$$2R_1 + R_2 \leq A'_1 + G'_1 + E'_2 \quad (28\text{-}8)$$
$$2R_2 + R_1 \leq A'_2 + G'_2 + E'_1 \quad (28\text{-}9),$$

where $A'_i$ , $D'_i$ , $E'_i$ , $G'_i$ , $i = 1{,}2$ are the same as in Theorems 18.

**Proof.** The proof is the same as the proof for Theorem 9 and hence, is omitted.

**An easily comparable form for the Hodtani-CMG ($\boldsymbol{R_1}$ , $\boldsymbol{R_2}$) region**

**Theorem 21.** Explanatory and easily comparable form of $\mathcal{R}_{H0d-CMG}$ in Theorem 20 can be described as $\mathcal{R}_{Hod-CMG-equi}$. (comparable to $\mathcal{R}_{Hod}^{mod}$. in Theorem 17) satisfying thirteen relations for $(R_1 ,\ R_2)$.

$$R_1 \leq D'_1 \quad (29\text{-}1)$$
$$\boldsymbol{R_1 \leq A'_1 + E'_2} \quad (29\text{-}2)$$
$$R_2 \leq D'_2 \quad (29\text{-}3)$$
$$\boldsymbol{R_2 \leq A'_2 + E'_1} \quad (29\text{-}4)$$
$$R_1 + R_2 \leq A'_1 + G'_2 \quad (29\text{-}5)$$
$$R_1 + R_2 \leq A'_2 + G'_1 \quad (29\text{-}6)$$
$$R_1 + R_2 \leq E'_1 + E'_2 \quad (29\text{-}7)$$
$$2R_1 + R_2 \leq A'_1 + G'_1 + E'_2 \quad (29\text{-}8)$$
$$2R_2 + R_1 \leq A'_2 + G'_2 + E'_1 \quad (29\text{-}9)$$
$$\left\{\begin{array}{ll} R_1 \leq A'_1 + F'_2 & (29\text{-}10) \\ R_2 \leq A'_2 + F'_1 & (29\text{-}11) \\ 2R_1 + R_2 \leq 2A'_1 + F'_2 + E'_2 & (29\text{-}12) \\ 2R_2 + R_1 \leq 2A'_2 + E'_1 + F'_1 & (29\text{-}13) \end{array}\right\}$$

where $A'_i$ , $D'_i$ , $E'_i$ , $G'_i$ , $i = 1{,}2$ are the same as in Theorems 18.

**Proof.** In view of easily provable relation $E'_i \leq G'_i = F'_i$ and the inequalities in Theorem 20, (29-10) and (29-11) are redundant due to (29-2) and (29-4), respectively ; and also, (29-2,5) and (29-4,6) yield (29-12) and (29-13), respectively.

## VIII. COMPARISON OF THE REGIONS

In this important section, first we prove a basic theorem (Theorem 22) and then, compare our region with the HK region, the HK-CMG or the so called CMG region and newly defined Hodtani-CMG region, as follows.

### A. An exemplifying theorem for comparison purposes of the regions

First of all, based on the Wyner common information and on the latest studies on common information [42] as explained and exemplified in section III-a , we consider a simple dependency structure between input auxiliary random variables and prove a basic theorem for comparing the regions.

**Theorem 22 [exemplifying theorem, accompanied with the equalities in Theorems 19 and 7].** Based on the Wyner common information (as in [42] and explanations in section III-a), by considering message random variables in the Hodtani region and distribution (16) as $U_{1d} = (U_1, K_1)$, $W_{1d} = (W_1, K_1)$, $W_{2d} = (W_2, K_2)$, $U_{2d} = (U_2, K_2)$, with all independent variables $U_1, W_1, W_2, U_2$ (the HK and the CMG message variables), $K_1, K_2$ (common information random variables), we have:

(22-I) $A_i = A'_i$ , $D_i = D'_i = B'_i$ , $E_i = E'_i$ , $G_i = G'_i = F'_i$

(22-II) $a_i = a'_i$ , $d_i = d'_i = b'_i$ , $e_i = e'_i$ , $g_i = g'_i = f'_i$ and

(22-III)

$$A_i \geq a_i \quad (30\text{-}1)$$
$$B_i \geq b_i \quad (30\text{-}2)$$
$$C_i \geq c_i \quad (30\text{-}3)$$
$$D_i \geq d_i \quad (30\text{-}4)$$
$$E_i \geq e_i \quad (30\text{-}5)$$
$$F_i \geq f_i \quad (30\text{-}6)$$
$$G_i \geq g_i \quad (30\text{-}7),$$

where, $i = 1,2$ and $A_i = A'_i, B_i, C_i, D_i = D'_i, E_i = E'_i, F_i, G_i = G'_i$ (Theorems 14.,18, 19) are the bound constants in the Hodtani and the Hodtani-CMG regions, and $a_i = a'_i$ , $b_i$, $c_i$ , $d_i = d'_i$ , $e_i = e'_i$ , $f_i$ , $g_i = g'_i$ (Theorems 1,6, 7) are the bound constants in the HK and the CMG regions.
And the inequalities in (30-1,..,7) turn out to be the equalities for independent situation of random variables or for $K_1 = K_2 = \emptyset$ .

**Proof.** The equalities in (22-I) and (22-II) have been proved in Theorems 19 and 7, respectively. Therefore, it is sufficient to prove that the inequalities (30-1,…,7) hold. Due to the independency of variables $U_i, W_i, K_i$ , $i = 1,2$, the inequality $I(X;Y|Z) \geq I(X;Y)$ for independent $(X,Z)$, and other information theory inequalities, the proof is done easily as follows (for $i = 1$).

- $A_1 = I(Y_1;\, U_{1d}|W_{1d}W_{2d}Q) = I(Y_1;\, U_1K_1|W_1K_1W_2K_2Q) = I(Y_1;\, U_1|W_1K_1K_2W_2Q) \geq$
  $I(Y_1;\, U_1|W_1W_2Q) = a_1 = a'_1$
- $B_1 = I(Y_1;\, W_{1d}|U_{1d}W_{2d}Q) + I(U_{1d}; W_{1d}|Q) = I(Y_1;\, W_1|U_1K_1K_2W_2Q) + H(K_1|Q) \geq$
  $I(Y_1;\, W_1|U_1W_2Q) + H(K_1|Q) = b_1 + H(K_1|Q) \Rightarrow B_1 \geq b_1$
- $C_1 = I(Y_1;\, W_{2d}|U_{1d}W_{1d}Q) + I(U_{1d}; W_{1d}|Q) = I(Y_1;\, W_2K_2|U_1W_1K_1Q) + H(K_1|Q) =$
  $I(Y_1;\, W_2|U_1W_1K_1Q) + I(Y_1; K_2|U_1W_1K_1Q) + H(K_1|Q) \geq$
  $c_1 + I(Y_1; K_2|U_1W_1K_1Q) + H(K_1|Q) \Rightarrow\ C_1 \geq c_1 + I(Y_1; K_2|U_1W_1K_1Q) + H(K_1|Q)\ \Rightarrow$
  $C_1 \geq c_1$
- $D_1 = I(Y_1;\, U_{1d}W_{1d}|W_{2d}Q) = I(Y_1;\, U_1W_1K_1|W_2K_2Q) =$
  $I(Y_1;\, U_1W_1|W_2K_2Q) + I(Y_1; K_1|QW_2K_2U_1W_1) \geq$
  $I(Y_1;\, U_1W_1|QW_2) + I(Y_1; K_1|QW_2K_2U_1W_1) = d_1 + I(Y_1; K_1|QW_2K_2U_1W_1) \Rightarrow D_1 \geq d_1 = d'_1$
- $E_1 = I(Y_1;\, U_{1d}W_{2d}|QW_{1d}) = I(Y_1;\, U_1W_2K_1K_2|QW_1K_1) = I(Y_1;\, U_1W_2K_2|QW_1K_1) =$
  $I(Y_1; K_2|QW_1K_1) + I(Y_1;\, U_1W_2|QW_1K_1K_2) \geq I(Y_1; K_2|QW_1K_1) + I(Y_1; U_1W_2|QW_1) \geq$
  $I(Y_1; K_2|QW_1K_1) + e_1 \Rightarrow E_1 \geq e_1 = e'_1$
- $F_1 = I(Y_1;\, W_{1d}W_{2d}|QU_{1d}) + I(U_{1d}; W_{1d}|Q) = I(Y_1; W_1W_2K_2|QU_1K_1) + H(K_1|Q) \geq I(Y_1; W_1W_2|QU_1) =$
  $f_1 \Rightarrow F_1 \geq f_1$
- $G_1 = I(Y_1; U_{1d}W_{1d}W_{2d}|Q) = I(Y_1;\, U_1W_1W_2K_1K_2|Q) \geq$
  $I(Y_1; U_1W_1W_2|Q) = g_1\ \Rightarrow G_1 \geq g_1 = g'_1$ .

The proof for $i = 2$ is the same as the above.

For the case of independent variables $(K_1 = K_2 = \emptyset)$ , all of the above inequalities obviously turn out to be equalities.

Therefore, the proof is completed.

**B. Comparisons**

In the previous section, we had the equivalency of the HK region to the HK-CMG (the CMG) region. Hence, the comparisons of the HK region with the Hodtani region, the HK-CMG or the CMG region to the Hodtani-CMG region and the Hodtani region to the Hodtani-CMG region suffice.

We show that the previous known regions (the HK and the CMG regions) are special cases of our new region (the Hodtani and the Hodtani-CMG regions) as follows.

- **First Comparison: The HK region is included in the Hodtani region**

**Input distributions and coding strategies**

The input distribution in the HK coding is (2), where the message variables are independent.
The HK region has been obtained using superposition coding-jointly decoding.

In the Hodtani region, (16) is input distribution where the message variable are dependent and the region has been obtained by random binning technique-jointly decoding.

**Comparing the HK and Hodtani regions**

It is sufficient to compare the Hodtani region to the HK region, by reviewing $S_{IC}^{HK}$ in Theorem 1 and $S_{IC}^{Hod}$ in Theorem 14; $\mathcal{R}_{HK}$ in Theorem 2 or 3 and $\mathcal{R}_{Hod}$ in Theorem 16, and $\mathcal{R}_{HK}^{mod.}$ in Theorem 5 and $\mathcal{R}_{Hod}^{mod.}$ in Theorem 17, thereby showing that the HK region is included in the Hodtani region:

**Theorem 23.** $S_{IC}^{HK}$ in Theorem 1 is a special case of $S_{IC}^{Hod}$ in Theorem 14 or

$$(T_1, S_1, T_2, S_2) \in S_{IC}^{HK} \Longrightarrow (T_1, S_1, T_2, S_2) \in S_{IC}^{Hod} \ ,$$

and the inverse is not true.

**Proof.** Due to Theorem 22 (22-III), the proof is obvious.

Note that in the case of independency of $U_i\ , W_i$ given $Q\ \ ,\ i = 1{,}2$, all of terms of $S_{IC}^{Hod}$ turn out to be equal to the associated terms in $S_{IC}^{HK}$.

**Theorem 24.** $\mathcal{R}_{HK}$ = (4-1,…,9) in Theorem 2 or (4-1,…,13) in Theorem 3 is a special case of $\mathcal{R}_{Hod}$ = (18-1,…,13) in Theorem 16 or:

$$(R_1\ , R_2) \in\ \mathcal{R}_{HK} \Longrightarrow (R_1\ , R_2) \in \mathcal{R}_{Hod}\ ,$$

and the inverse is not true.

**Proof.** Due to Theorem 22 (22-III), the proof is obvious.

As explained in the outline of the proof for Theorem 16, in the case of independency of $U_i\ , W_i$ given $Q\ \ ,\ i = 1{,}2$, $\mathcal{R}_{Hod}$ = (18-1,…,13) is reduced to the $\mathcal{R}_{HK}$ = (4-1,…,9).

**Theorem 25.** $\mathcal{R}_{HK}^{mod.}$ (8-1,…,13) in Theorem 5 is a special case of $\mathcal{R}_{Hod}^{mod.}$ (19-1,…,13) in Theorem 17 or

$$(R_1\ , R_2) \in\ \mathcal{R}_{HK}^{mod.} \Longrightarrow (R_1\ , R_2) \in \mathcal{R}_{Hod}^{mod.}\ ,$$

and the inverse is not true.

**Proof.** Due to Theorem 22 (22-III), the proof is obvious.

- **Second Comparison: In accordance with Theorem 13, the CMG region is equivalent to the HK region.**
- **Third Comparison: As a result of first and second comparisons, the CMG and the mod-CMG regions like the HK region are included in the Hodtani region.**
- **Fourth Comparison: The HK-CMG region or the so called CMG region is included in the Hodtani-CMG region**

We explained the CMG region in detail (section V ), and defined the HK-CMG region and the Hodtani-CMG region in section VII. Now, we compare the two regions.

**Input distributions and coding strategies**

The input distribution in the HK-CMG coding is (9-c), where the message variables are independent.

The HK-CMG region has been obtained using superposition coding-jointly decoding.

In the Hodtani-CMG region, (23) is the input distribution where the message variables are dependent and the region has been obtained by superposition coding-jointly decoding.

**Comparing the HK-CMG (or CMG) and the Hodtani-CMG regions**

**Theorem 26.** $S_{IC}^{CMG}$ in Theorem 6 is a special case of $S_{IC}^{Hod-CMG}$ in Theorem 18 or

$$(T_1, S_1, T_2, S_2) \in S_{IC}^{CMG} \Longrightarrow (T_1, S_1, T_2, S_2) \in S_{IC}^{Hod-CMG}\ ,$$

and the inverse is not true.

**Proof.** Due to the result of Theorem 22 ($a'_i \leq A'_i, i = 1{,}2\ , etc.$), the proof is obvious.

**Theorem 27.** $\mathcal{R}_{CMG}$ in Theorem 9 is a special case of $\mathcal{R}_{H0d-CMG}$ in Theorem 20 or

$$(R_1\ , R_2) \in\ \mathcal{R}_{CMG} \Longrightarrow (R_1\ , R_2) \in \mathcal{R}_{Hod-CMG}\ ,$$

**Proof.** Due to the inequalities (30-1,4,5,7) in Theorem 22, the proof is obvious.

- **Fifth Comparison: The Hodtani region is equivalent to the Hodtani-CMG region**

**Theorem 28.** $\mathcal{R}_{Hod}^{mod.}$ in Theorem 17 is equivalent to $\mathcal{R}_{Hod-CMG-equi.}$ in Theorem 21or

$$(R_1\ , R_2) \in\ \mathcal{R}_{Hod}^{mod.} \Leftrightarrow (R_1\ , R_2) \in\ \mathcal{R}_{Hod-CMG-equi.},$$

**Proof.** Due to $A_i = A'_i, D_i = D'_i, E_i = E'_i, G_i = G'_i$ in Theorem 22 (22-I), the proof is obvious.

- **Sixth Comparison: As a result of the above comparisons, the previously known equivalent regions (the HK and the CMG regions) are included in new equivalent regions ( the Hodtani and the Hodtani-CMG regions).**

## IX. CONCLUSION

By making two novel changes in the HK coding, i.e., allowing the input auxiliary random variables to be correlated and using binning scheme and jointly decoding, first, we obtained an improved version of the HK region for the general IC, as the Hodtani region. And, second, we obtained the Hodtani-CMG region as an improved version of the CMG region. We have shown that the previous regions (the HK and the CMG regions) are special cases of our new region (the Hodtani and the Hodtani-CMG regions). In our region, interestingly, every rate for the IC, has generally three terms: the first is a general HK term with dependent variables, the second is due to the input correlation and the third is a result of binning scheme.

## APPENDIX A

**The proof of Theorem 14**

It is sufficient to show that any element of $S_{IC}^{Hod}(Z_2)$ for each $Z_2 \in \mathcal{P}_{IC}^{Hod}$ is achievable. So, fix $Z_2 = (QU_{1d}W_{1d}U_{2d}W_{2d}X_1X_2Y_1Y_2)$ and take any $(T_1, S_1, T_2, S_2)$ satisfying the constraints of the theorem.

**Codebook generation:** Consider $n > 0$ , some distribution of the form (16) and

$p(u_{1d}|q) = \sum_{w_{1d}} p(w_{1d}|q)\, p(u_{1d}|qw_{1d})$

$p(u_{2d}|q) = \sum_{w_{2d}} p(w_{2d}|q)\, p(u_{2d}|qw_{2d})$.

Therefore, by using binning scheme we can generate sequences of $\boldsymbol{u_{1d}}$ and $\boldsymbol{u_{2d}}$ independently of $\boldsymbol{w_{1d}}$ and $\boldsymbol{w_{2d}}$. So,

1. generate a n-sequence $\boldsymbol{q}$, i.i.d. according to $\prod_{i=1}^{n} p(q_i)$, and then, for the codeword $\boldsymbol{q}$:
2. Generate $\lfloor 2^{nT_1} \rfloor$ conditionally independent codewords $\boldsymbol{w_{1d}}(j),\ j \in \{1,2,\cdots,\lfloor 2^{nT_1} \rfloor\}$ according to $\prod_{i=1}^{n} p(w_{1di}|q_i)$.
3. Generate $\lfloor 2^{ns_1} \rfloor$ (**$\boldsymbol{s_1}$ is small letter**) n-sequence $\boldsymbol{u_{1d}}(l),\ l \in \{1,\cdots,\lfloor 2^{ns_1} \rfloor\}$, i.i.d. according to $\prod_{i=1}^{n} p(u_{1di}|q_i)$ and throw them randomly into $\lfloor 2^{nS_1} \rfloor$ (**$\boldsymbol{S_1}$ is capital letter**) bins such that the sequence $\boldsymbol{u_{1d}}(l)$ in bin $b_1$ is denoted as $\boldsymbol{u_{1d}}(b_1, l),\ b_1 \in \{1,\cdots,\lfloor 2^{nS_1} \rfloor\}$.
4. Generate $\lfloor 2^{nT_2} \rfloor$ n-sequence $\boldsymbol{w_{2d}}(m),\ m \in \{1,\cdots,\lfloor 2^{nT_2} \rfloor\}$, i.i.d. according to $\prod_{i=1}^{n} p(w_{2di}|q_i)$.
5. Generate $\lfloor 2^{ns_2} \rfloor$ (**$\boldsymbol{s_2}$ is small letter**) n-sequence $\boldsymbol{u_{2d}}(k),\ k \in \{1,\cdots,\lfloor 2^{ns_2} \rfloor\}$, i.i.d. according to $\prod_{i=1}^{n} p(u_{2di}|q_i)$ and throw them randomly into $\lfloor 2^{nS_2} \rfloor$ (**$\boldsymbol{S_2}$ is capital letter**) bins such that the sequence $\boldsymbol{u_{2d}}(k)$ in bin $b_2$ is denoted as $\boldsymbol{u_{2d}}(b_2, k),\ b_2 \in \{1,\cdots,\lfloor 2^{nS_2} \rfloor\}$.

**Encoding:** The aim is to send a two dimensional message at each sender. The messages are mapped into $\boldsymbol{x_1}$ and $\boldsymbol{x_2}$ through deterministic encoding functions $f_1$ and $f_2$ (as in [6]). The sender $TX_1$ to send $(j, b_1)$, knowing $\boldsymbol{q}$ looks for $\boldsymbol{w_{1d}}(j)$ and finds a sequence $\boldsymbol{u_{1d}}(b_1, l)$ in bin $b_1$ such that $(\boldsymbol{q}, \boldsymbol{w_{1d}}(j), \boldsymbol{u_{1d}}(b_1, l)) \in A_\varepsilon^n$ ; then generates $\boldsymbol{x_1}$ i.i.d. according to $x_{1i} = f_1(w_{1di}(j), u_{1di}(b_1, l)|q_i)$ , $i = 1,\cdots,n$; and sends it. The sender $TX_2$ to send $(m, b_2)$, knowing $\boldsymbol{q}$ looks for $\boldsymbol{w_{2d}}(m)$ and finds a sequence $\boldsymbol{u_{2d}}(k)$ in bin $b_2$ such that $(\boldsymbol{q}, \boldsymbol{w_{2d}}(m), \boldsymbol{u_{2d}}(b_2, k)) \in A_\varepsilon^n$ ; then generates $\boldsymbol{x_2}$ i.i.d. according to $x_{2i} = f_2(w_{2di}(m), u_{2di}(b_2, k)|q_i)$ , $i = 1,\cdots,n$ and sends it.

**Decoding and analysis of error probability:** The receivers $RX_1$ and $RX_2$ decode the corresponding messages, based on strong joint typicality [6]. It is assumed that all messages are equiprobable. Without loss of generality, we may confine ourselves to the situation where $(j = 1, b_1 = 1; m = 1, b_2 = 1)$ was sent.

The receiver $RX_1$, by receiving $\boldsymbol{y_1}$ and knowing $\boldsymbol{q}$, decodes $j = 1, b_1 = 1, m = 1$ or $j(b_1, l)\, m = 1(1, l)1$ simultaneously [6]. We define the event $E_{j\,(b_1,l)\,m}$ and $P_{e_1}^n$ as follows.

$E_{j\,(b_1,l)m} = \{(\boldsymbol{q}, \boldsymbol{w_{1d}}(j), \boldsymbol{u_{1d}}(b_1, l), \boldsymbol{w_{2d}}(m), \boldsymbol{y_1}) \in A_\varepsilon^n\}$

$P_{e_1}^{(n)} = P\{E_{1(1,l)1}^C \cup E_{j\,(b_1,l)m \neq 1(1,l)1}\} \leq P(E_{1(1,l)1}^C) + \sum_{j\,(b_1,l)m \neq 1(1,l)1} P(E_{j\,(b_1,l)m}) \leq \varepsilon + \sum_{j \neq 1, b_1 = m = 1}^{\boxed{1}} \cdots + \sum_{b_1 \neq 1, j = m = 1}^{\boxed{2}} \cdots + \sum_{m \neq 1, j = b_1 = 1}^{\boxed{3}} \cdots + \sum_{j \neq 1, m \neq 1, b_1 = 1}^{\boxed{4}} \cdots + \sum_{j \neq 1, b_1 \neq 1, m = 1}^{\boxed{5}} \cdots + \sum_{m \neq 1, b_1 \neq 1, j = 1}^{\boxed{6}} \cdots + \sum_{j \neq 1, m \neq 1, b_1 \neq 1}^{\boxed{7}} \cdots$

In accordance with the codebook generation and the original distribution (16), we prove the Theorem.

Note that the **proof** is done directly in view of **how** the sequences have been generated, the corresponding $\varepsilon-$typicallity **probabilities** and the **distribution (16**) for random variables $U_{1d}, W_{1d}, U_{2d}, W_{2d}, Q$ .

- $\sum_{j \neq 1, b_1 = m = 1}^{\boxed{1}} p(E_{j\,(1,l)1}) \leq 2^{nT_1} \cdot p((\boldsymbol{q}, \boldsymbol{w_{1d}}(j), \boldsymbol{u_{1d}}(1, l), \boldsymbol{w_{2d}}(1), \boldsymbol{y_1}) \in A_\varepsilon^n) \leq$

$$2^{nT_1} \sum_{(\boldsymbol{q}, \boldsymbol{w_{1d}}(j), \boldsymbol{u_{1d}}(1,l), \boldsymbol{w_{2d}}(1), \boldsymbol{y_1}) \in A_\varepsilon^n} p(\boldsymbol{q}, \boldsymbol{w_{1d}}(j), \boldsymbol{u_{1d}}(1, l), \boldsymbol{w_{2d}}(1), \boldsymbol{y_1}) \leq$$

$$2^{nT_1} \cdot 2^{nH(QW_{1d}U_{1d}W_{2d}Y_1)} . p(\boldsymbol{q}) p(\boldsymbol{w_{1d}}|\boldsymbol{q}) p(\boldsymbol{w_{2d}}|\boldsymbol{q}) p(\boldsymbol{u_{1d}}|\boldsymbol{q}) p(\boldsymbol{y_1}|\boldsymbol{q}\boldsymbol{u_{1d}}\boldsymbol{w_{2d}}) \leq$$

$$2^{nT_1} \cdot 2^{nH(QW_{1d}U_{1d}W_{2d}Y_1)} \cdot 2^{-nH(Q)} \cdot 2^{-nH(W_{1d}|Q)} \cdot 2^{-nH(W_{2d}|Q)} \cdot 2^{-nH(U_{1d}|Q)} \cdot 2^{-nH(Y_1|QU_{1d}W_{2d})} \stackrel{a}{=}$$

$$2^{nT_1} \cdot 2^{nH(U_{1d}Y_1|QW_{1d}W_{2d}) - nH(U_{1d}|Q) - nH(Y_1|QU_{1d}W_{2d})} \stackrel{b}{=}$$

$$2^{nT_1 + nH(U_{1d}|QW_{1d}W_{2d}) + nH(Y_1|QW_{1d}W_{2d}U_{1d}) - nH(U_{1d}|Q) - nH(Y_1|QU_{1d}W_{2d})} =$$

$$2^{nT_1 - nI(U_{1d};W_{1d}|Q) - n(H(Y_1|QU_{1d}W_{2d}) - H(Y_1|QW_{1d}W_{2d}U_{1d}))} = 2^{nT_1 - nI(U_{1d};W_{1d}|Q) - nI(Y_1;W_{1d}|QU_{1d}W_{2d})},$$

where, a (b) follows from the independence of $W_2(U_1)$ from $W_1(W_2)$ given $Q(QW_1)$ respectively. These independences are also used in the following $\boxed{2} - \boxed{7}$.

- $\sum_{b_1 \neq 1, j=m=1}^{\boxed{2}} p\left((\boldsymbol{q}, \boldsymbol{w_{1d}}(1), \boldsymbol{u_{1d}}(b_1, l), \boldsymbol{w_{2d}}(1), \boldsymbol{y_1}) \in A_\varepsilon^n\right) \leq$

  $2^{ns_1} p\left((\boldsymbol{q}, \boldsymbol{w_{1d}}(1), \boldsymbol{u_{1d}}(b_1, l), \boldsymbol{w_{2d}}(1), \boldsymbol{y_1}) \in A_\varepsilon^n\right) \leq$

  $2^{ns_1} \cdot \sum_{(\boldsymbol{q}, \boldsymbol{w_{1d}}(1), \boldsymbol{u_{1d}}(b_1, l), \boldsymbol{w_{2d}}(1), \boldsymbol{y_1}) \in A_\varepsilon^n} p(\boldsymbol{q}, \boldsymbol{w_{1d}}(1), \boldsymbol{u_{1d}}(b_1, l), \boldsymbol{w_{2d}}(1), \boldsymbol{y_1}) \leq$

  $2^{ns_1} \cdot 2^{nH(QW_{1d}U_{1d}W_{2d}Y_1)} \cdot p(\boldsymbol{q}) p(\boldsymbol{w_{1d}}|\boldsymbol{q}) p(\boldsymbol{w_{2d}}|\boldsymbol{q}) p(\boldsymbol{u_{1d}}|\boldsymbol{q}) p(\boldsymbol{y_1}|\boldsymbol{q}\boldsymbol{w_{1d}}\boldsymbol{w_{2d}}) \leq$

  $2^{ns_1} \cdot 2^{nH(QW_{1d}U_{1d}W_{2d}Y_1)} \cdot 2^{-nH(Q)} \cdot 2^{-nH(W_{1d}|Q)} \cdot 2^{-nH(W_{2d}|Q)} \cdot 2^{-nH(U_{1d}|Q)} \cdot 2^{-nH(Y_1|QW_{1d}W_{2d})} =$

  $2^{ns_1} \cdot 2^{nH(QW_{1d}U_{1d}W_{2d}Y_1) - nH(QW_{1d}W_{2d}) - nH(U_{1d}|Q) - nH(Y_1|QW_{1d}W_{2d})} =$

  $2^{ns_1} \cdot 2^{nH(U_{1d}Y_1|QW_{1d}W_{2d}) - nH(U_{1d}|Q) - nH(Y_1|QW_{1d}W_{2d})} =$

  $2^{ns_1} \cdot 2^{nH(U_{1d}|QW_{1d}W_{2d}) + nH(Y_1|QW_{1d}W_{2d}U_{1d}) - nH(U_{1d}|Q) - nH(Y_1|QW_{1d}W_{2d})} =$

  $2^{ns_1} \cdot 2^{nH(U_{1d}|QW_{1d}) - nH(U_{1d}|Q) + nH(Y_1|QW_{1d}W_{2d}U_1) - nH(Y_1|QW_{1d}W_{2d})} =$

  $2^{ns_1} \cdot 2^{-n\left(I(U_{1d}; W_{1d}|Q) + I(Y_1; U_{1d}|QW_{1d}W_{2d})\right)}$

- $\sum_{m \neq 1, j=b_1=1}^{\boxed{3}} p\left((\boldsymbol{q}, \boldsymbol{w_{1d}}(1), \boldsymbol{u_{1d}}(1, l), \boldsymbol{w_{2d}}(m), \boldsymbol{y_1}) \in A_\varepsilon^n\right) \leq$

  $2^{nT_2} \cdot \sum_{(\boldsymbol{q}, \boldsymbol{w_{1d}}(1), \boldsymbol{u_{1d}}(1, l), \boldsymbol{w_{2d}}(m), \boldsymbol{y_1}) \in A_\varepsilon^n} p(\boldsymbol{q}, \boldsymbol{w_{1d}}(1), \boldsymbol{u_{1d}}(1, l), \boldsymbol{w_{2d}}(m), \boldsymbol{y_1}) \leq$

  $2^{nT_2} \cdot 2^{nH(QW_{1d}U_{1d}W_{2d}Y_1)} \cdot p(\boldsymbol{q}) p(\boldsymbol{w_{1d}}|\boldsymbol{q}) p(\boldsymbol{w_{2d}}|\boldsymbol{q}) p(\boldsymbol{u_{1d}}|\boldsymbol{q}) p(\boldsymbol{y_1}|\boldsymbol{q}\boldsymbol{w_{1d}}\boldsymbol{u_{1d}}) \leq$

  $2^{nT_2} \cdot 2^{nH(QW_{1d}U_{1d}W_{2d}Y_1) - nH(QW_{1d}W_{2d}) - nH(U_{1d}|Q) - nH(Y_1|QW_{1d}U_{1d})} =$

  $2^{nT_2} \cdot 2^{nH(U_{1d}Y_1|QW_{1d}W_{2d}) - nH(U_{1d}|Q) - nH(Y_1|QW_{1d}U_{1d})} =$

  $2^{nT_2} \cdot 2^{nH(U_{1d}|QW_{1d}W_{2d}) - nH(U_{1d}|Q) + nH(Y_1|QW_{1d}W_{2d}U_1) - nH(Y_1|QW_{1d}U_{1d})} =$

  $2^{nT_2} \cdot 2^{-n\left(I(U_{1d}; W_{1d}|Q) - I(Y_1; W_{2d}|QW_{1d}U_{1d})\right)}$

- $\sum_{j \neq 1, m \neq 1, b_1=1}^{\boxed{4}} p\left((\boldsymbol{q}, \boldsymbol{w_{1d}}(j), \boldsymbol{u_{1d}}(1, l), \boldsymbol{w_{2d}}(m), \boldsymbol{y_1}) \in A_\varepsilon^n\right) \leq$

  $2^{n(T_1+T_2)} \cdot 2^{nH(QW_{1d}U_{1d}W_{2d}Y_1)} \cdot p(\boldsymbol{q}) p(\boldsymbol{w_{1d}}|\boldsymbol{q}) p(\boldsymbol{w_{2d}}|\boldsymbol{q}) p(\boldsymbol{u_{1d}}|\boldsymbol{q}) p(\boldsymbol{y_1}|\boldsymbol{q}\boldsymbol{u_{1d}}) \leq$

  $2^{n(T_1+T_2) + nH(QW_{1d}U_{1d}W_{2d}Y_1) - nH(QW_{1d}W_{2d}) - nH(U_{1d}|Q) - nH(Y_1|QU_{1d})} =$

  $2^{n(T_1+T_2) - n\left(I(U_{1d}; W_{1d}|Q) + I(Y_1; W_{1d}W_{2d}|QU_{1d})\right)}$

- $\sum_{j \neq 1, b_1 \neq 1, m=1}^{\boxed{5}} p\left((\boldsymbol{q}, \boldsymbol{w_{1d}}(j), \boldsymbol{u_{1d}}(b_1, l), \boldsymbol{w_{2d}}(1), \boldsymbol{y_1}) \in A_\varepsilon^n\right) \leq$

  $2^{n(T_1+s_1)} \cdot p(\boldsymbol{q}, \boldsymbol{w_{1d}}(j), \boldsymbol{u_{1d}}(b_1, l), \boldsymbol{w_{2d}}(1), \boldsymbol{y_1}) \leq$

  $2^{n(T_1+s_1)} \cdot 2^{nH(QW_{1d}W_{2d}U_{1d}Y_1)} \cdot p(\boldsymbol{q}) p(\boldsymbol{w_{1d}}|\boldsymbol{q}) p(\boldsymbol{w_{2d}}|\boldsymbol{q}) p(\boldsymbol{u_{1d}}|\boldsymbol{q}) p(\boldsymbol{y_1}|\boldsymbol{q}\boldsymbol{w_{2d}}) \leq$

  $2^{n(T_1+s_1)} \cdot 2^{nH(QW_{1d}W_{2d}U_{1d}Y_1) - nH(QW_{1d}W_{2d}) - nH(U_{1d}|Q) - nH(Y_1|QW_{2d})} =$

  $2^{n(T_1+s_1)} \cdot 2^{nH(U_{1d}Y_1|QW_{1d}W_{2d}) - nH(U_{1d}|Q) - nH(Y_1|QW_{2d})} =$

  $2^{n(T_1+s_1)} \cdot 2^{nH(U_{1d}|QW_{1d}W_{2d}) - nH(U_{1d}|Q) + nH(Y_1|QW_{1d}W_{2d}U_1) - nH(Y_1|QW_{2d})} =$

  $2^{n(T_1+s_1)} \cdot 2^{-n\left(I(U_{1d}; W_{1d}|Q) + I(Y_1; U_{1d}W_{1d}|QW_{2d})\right)}$

- $\sum_{m\neq 1,b_1\neq 1,j=1}^{\boxed{6}} p\big((\boldsymbol{q},\boldsymbol{w_{1d}}(1),\boldsymbol{u_{1d}}(b_1,l),\boldsymbol{w_{2d}}(m),\boldsymbol{y_1})\in A_\varepsilon^n\big)\leq$

  $2^{n(T_2+s_1)}\cdot p(\boldsymbol{q},\boldsymbol{w_{1d}}(1),\boldsymbol{u_{1d}}(b_1,l),\boldsymbol{w_{2d}}(m),\boldsymbol{y_1})\leq$

  $2^{n(T_2+s_1)}\cdot 2^{nH(QW_{1d}W_{2d}U_{1d}Y_1)}\cdot p(\boldsymbol{q})p(\boldsymbol{w_{1d}}|\boldsymbol{q})p(\boldsymbol{w_{2d}}|\boldsymbol{q})p(\boldsymbol{u_{1d}}|\boldsymbol{q})p(\boldsymbol{y_1}|\boldsymbol{q}\boldsymbol{w_{1d}})\leq$

  $2^{n(T_2+s_1)}\cdot 2^{nH(QW_{1d}W_{2d}U_{1d}Y_1)-nH(QW_{1d}W_{2d})-nH(U_{1d}|Q)-nH(Y_1|QW_{1d})}=$

  $2^{n(T_2+s_1)}\cdot 2^{-n(I(U_{1d};W_{1d}|Q)+I(Y_1;U_{1d}W_{2d}|QW_{1d}))}$

- $\sum_{j\neq 1,m\neq 1,b_1\neq 1}^{\boxed{7}} p\big((\boldsymbol{q},\boldsymbol{w_{1d}}(j),\boldsymbol{u_{1d}}(b_1,l),\boldsymbol{w_{2d}}(m),\boldsymbol{y_1})\in A_\varepsilon^n\big)\leq$

  $2^{n(T_1+T_2+s_1)}\cdot\sum_{(\boldsymbol{q},\boldsymbol{w_{1d}}(j),\boldsymbol{u_{1d}}(b_1,l),\boldsymbol{w_{2d}}(m),\boldsymbol{y_1})\in A_\varepsilon^n} p(\boldsymbol{q})p(\boldsymbol{w_{1d}}|\boldsymbol{q})p(\boldsymbol{w_{2d}}|\boldsymbol{q})p(\boldsymbol{u_{1d}}|\boldsymbol{q})p(\boldsymbol{y_1}|\boldsymbol{q})\leq$

  $2^{n(T_1+T_2+s_1)+nH(QW_{1d}W_{2d}U_{1d}Y_1)-nH(Q)-nH(W_{1d}|Q)-nH(W_{2d}|Q)-nH(U_{1d}|Q)-nH(Y_1|Q)}=$

  $2^{n(T_1+T_2+s_1)}\cdot 2^{-n(I(U_{1d};W_{1d}|Q)+I(Y_1;U_{1d}W_{2d}W_{1d}|Q))}$

Similarly, the other terms can be evaluated. In order to ($P_{e_1}^{(n)}\longrightarrow 0$ as the block length $n\longrightarrow\infty$), it is necessary and sufficient that:

$$\begin{cases} T_1\leq I(Y_1;W_{1d}|W_{2d}U_1Q)+I(U_{1d};W_{1d}|Q)\\ s_1\leq I(U_{1d};W_{1d}|Q)+I(Y_1;U_{1d}|QW_{1d}W_{2d})\\ T_2\leq I(U_{1d};W_{1d}|Q)+I(Y_1;W_{2d}|QW_{1d}U_{1d})\\ T_1+T_2\leq I(Y_1;W_{1d}W_{2d}|QU_{1d})+I(U_{1d};W_{1d}|Q)\\ s_1+T_1\leq I(Y_1;U_{1d}W_{1d}|QW_{2d})+I(U_{1d};W_{1d}|Q)\\ s_1+T_2\leq I(Y_1;U_{1d}W_{2d}|QW_{1d})+I(U_{1d};W_{1d}|Q)\\ s_1+T_1+T_2\leq I(Y_1;U_{1d}W_{1d}W_{2d}|Q)+I(U_{1d};W_{1d}|Q)\end{cases}\qquad (A_1),\ (s_1\text{ is small letter})$$

from where, considering the binning condition:
$I(U_{1d};W_{1d}|Q)\leq s_1-S_1$ or $S_1-s_1\leq -I(U_{1d};W_{1d}|Q)$,
the relations ($A_1$) yield to the constraints (17-1)-(17-7) in Theorem 14.

Error probability analysis for the receiver $RX_2$ can be done similarly and the inequalities (17-8)-(17-14) can be proved (for brevity, the details are omitted).

## APPENDIX B

**The proof of Theorem 16**

In Theorem 14, we set $S_i=R_i-T_i$ , $i=1,2$ in (17-1)-(17-14) and in $S_i\geq 0$ , $T_i\geq 0$ , $i=1,2$ .Then, in the first step, we collect the inequalities not including $T_1$ , also we collect the inequalities including $+T_1$ and including $-T_1$ and by adding these terms we determine the inequalities not including $T_1$ . Then, similarly in the second step we eliminate $T_2$ , ultimately we obtain 36 inequalities:

$$\begin{cases} -R_1\leq 0 & \text{(B-1)}\\ 0\leq C_1 & \text{(B-2)}\\ 0\leq E_1 & \text{(B-3)}\\ 0\leq B_2 & \text{(B-4)}\\ -R_2\leq 0 & \text{(B-5)}\\ 0\leq A_2 & \text{(B-6)}\end{cases}\qquad \begin{cases} R_1\leq D_1 & \text{(B-7)}\\ R_1\leq A_1+C_2 & \text{(B-8)}\\ R_1\leq A_1+F_2 & \text{(B-9)}\\ R_1\leq E_1+C_2 & \text{(B-10)}\\ R_1\leq G_1 & \text{(B-11)}\\ R_1\leq E_1+F_1 & \text{(B-12)}\\ R_1\leq E_1+F_2 & \text{(B-13)}\\ R_1\leq A_1+E_2 & \text{(B-14)}\end{cases}$$

$$\begin{cases} R_1 + R_2 \le A_1 + G_2 & \text{(B-19)} \\ R_1 + R_2 \le E_1 + E_2 & \text{(B-20)} \\ R_1 + R_2 \le E_1 + G_2 & \text{(B-21)} \\ R_1 + R_2 \le A_1 + F_2 + A_2 & \text{(B-22)} \\ R_1 + R_2 \le E_1 + C_2 + A_2 & \text{(B-23)} \\ R_1 + R_2 \le G_1 + A_2 & \text{(B-24)} \\ R_1 + R_2 \le A_1 + E_2 + C_1 & \text{(B-25)} \\ R_1 + R_2 \le A_1 + E_2 + E_1 & \text{(B-26)} \\ R_1 + R_2 \le A_1 + E_2 + B_2 & \text{(B-27)} \end{cases}$$

$$\begin{cases} R_2 \le D_2 & \text{(B-15)} \\ R_2 \le A_2 + C_1 & \text{(B-16)} \\ R_2 \le A_2 + E_1 & \text{(B-17)} \\ R_2 \le A_2 + B_2 & \text{(B-18)} \end{cases}$$

$$\begin{cases} R_1 + 2R_2 \le 2A_2 + E_1 + F_1 & \text{(B-28)} \\ R_1 + 2R_2 \le 2A_2 + E_1 + F_2 & \text{(B-29)} \\ R_1 + 2R_2 \le A_2 + E_1 + G_2 & \text{(B-30)} \end{cases}$$

$$\begin{cases} 2R_1 + R_2 \le 2A_1 + E_2 + F_2 & \text{(B-31)} \\ 2R_1 + R_2 \le A_1 + E_2 + E_1 + C_2 & \text{(B-32)} \\ 2R_1 + R_2 \le A_1 + E_2 + G_1 & \text{(B-33)} \end{cases}$$

$$\begin{cases} 3R_1 + 2R_2 \le 2A_1 + 2E_2 + E_1 + F_1 & \text{(B-34)} \\ 3R_1 + 2R_2 \le 2A_1 + 2E_2 + E_1 + F_2 & \text{(B-35)} \end{cases}$$

$$\{ 2R_1 + 2R_2 \le A_1 + E_1 + G_2 + E_2 \qquad \text{(B-36)}$$

We see that (B-1,…,6) are redundant. Also, by considering the inequalities in Theorem 15, we conclude the following:

(B-9,10,13), (B-11,12) are redundant due to (B-8), (B-7), respectively. (B-21,23,25,26), (B-22,27), (B-15) are redundant due to (B-20), (B-19), (B-15) respectively. Also, (B-32), (B-34), (B-35), (B-36) are redundant due to (B-20,8), (B-20,33), (B-20,19), respectively.

The remaining inequalities (B-7,8,14,15,16,17,19,20,24,28,30,31,33) constitute the 13 inequalities of Theorem 16 and the proof is completed.